 \documentclass[sigconf]{acmart}
 
\usepackage{booktabs}
\usepackage{balance}
\usepackage{multirow}
\usepackage{graphicx}
\usepackage{algorithm}
\usepackage{algpseudocode}
\usepackage{caption}
\usepackage[T1]{fontenc}
\usepackage{bm}
\usepackage{tabularx}
\usepackage{array} % 

\AtBeginDocument{%
  }

\copyrightyear{2025}
\acmYear{2025}
\setcopyright{cc}
\setcctype{by}
\acmConference[MM '25]{Proceedings of the 33rd ACM International Conference on Multimedia}{October 27--31, 2025}{Dublin, Ireland}
\acmBooktitle{Proceedings of the 33rd ACM International Conference on Multimedia (MM '25), October 27--31, 2025, Dublin, Ireland}
\acmDOI{10.1145/3746027.3754729}
\acmISBN{979-8-4007-2035-2/2025/10}

\begin{document}

\title{A Comprehensive Benchmark for Electrocardiogram Time-Series}

\author{Zhijiang Tang}
\authornote{Both authors contributed equally to this research.}
\email{tangzhijiang24@mails.ucas.ac.cn}
% \orcid{https://orcid.org/0000-0002-0560-4132}
\affiliation{%
  \institution{Hangzhou Institute for Advanced Study, University of Chinese Academy of Sciences}
  \city{Hangzhou}
  \state{Zhejiang}
  \country{China}
}

\author{Jiaxin Qi}
\authornotemark[1]
\email{jxqi@cnic.cn}
% \orcid{https://orcid.org/0000-0002-0560-4132}
\affiliation{%
  \institution{Computer Network Information Center, Chinese Academy of Sciences}
  \city{Beijing}
  \state{Beijing}
  \country{China}
}

\author{Yuhua Zheng}
\email{zhengyuhua@ucas.ac.cn}
% \orcid{https://orcid.org/0000-0002-0560-4132}
\affiliation{%
  \institution{Hangzhou Institute for Advanced Study, University of Chinese Academy of Sciences}
  \city{Hangzhou}
  \state{Zhejiang}
  \country{China}}

\author{Jianqiang Huang}
\authornote{Corresponding author.}
\email{jqhuang@cnic.cn}
\affiliation{%
  \institution{Computer Network Information Center, Chinese Academy of Sciences}
    \city{Beijing}
  \state{Beijing}
  \country{China}
}
\affiliation{%
  \institution{Hangzhou Institute for Advanced Study, University of Chinese Academy of Sciences}
  \city{Hangzhou}
  \state{Zhejiang}
  \country{China}
}

%%
%% Sometimes the addresses are too long to fit on the page.  In this
%% case uncomment the lines below and fill them accodingly.
%%
%% \authorsaddresses{Corresponding author: Ben Trovato,
%% \href{mailto:trovato@corporation.com}{trovato@corporation.com};
%% Institute for Clarity in Documentation, P.O. Box 1212, Dublin,
%% Ohio, USA, 43017-6221}

\begin{abstract}
Electrocardiogram~(ECG), a key bioelectrical time-series signal, is crucial for assessing cardiac health and diagnosing various diseases. Given its time-series format, ECG data is often incorporated into pre-training datasets for large-scale time-series model training. However, existing studies often overlook its unique characteristics and specialized downstream applications, which differ significantly from other time-series data, leading to an incomplete understanding of its properties.
In this paper, we present an in-depth investigation of ECG signals and establish a comprehensive benchmark, which includes (1) categorizing its downstream applications into four distinct evaluation tasks, (2) identifying limitations in traditional evaluation metrics for ECG analysis, and introducing a novel metric; (3) benchmarking state-of-the-art time-series models and proposing a new architecture.
Extensive experiments demonstrate that our proposed benchmark is comprehensive and robust. The results validate the effectiveness of the proposed metric and model architecture, which establish a solid foundation for advancing research in ECG signal analysis.
\end{abstract}

% \textbf{Code: }\href{https://github.com/ZhijiangTang/ECG-Benchmark}{https://github.com/ZhijiangTang/ECG-Benchmark}  



\maketitle

\section{Introduction}
\label{sec:intro}

\begin{figure}[htbp]
    \centering
        \includegraphics[width=\linewidth]{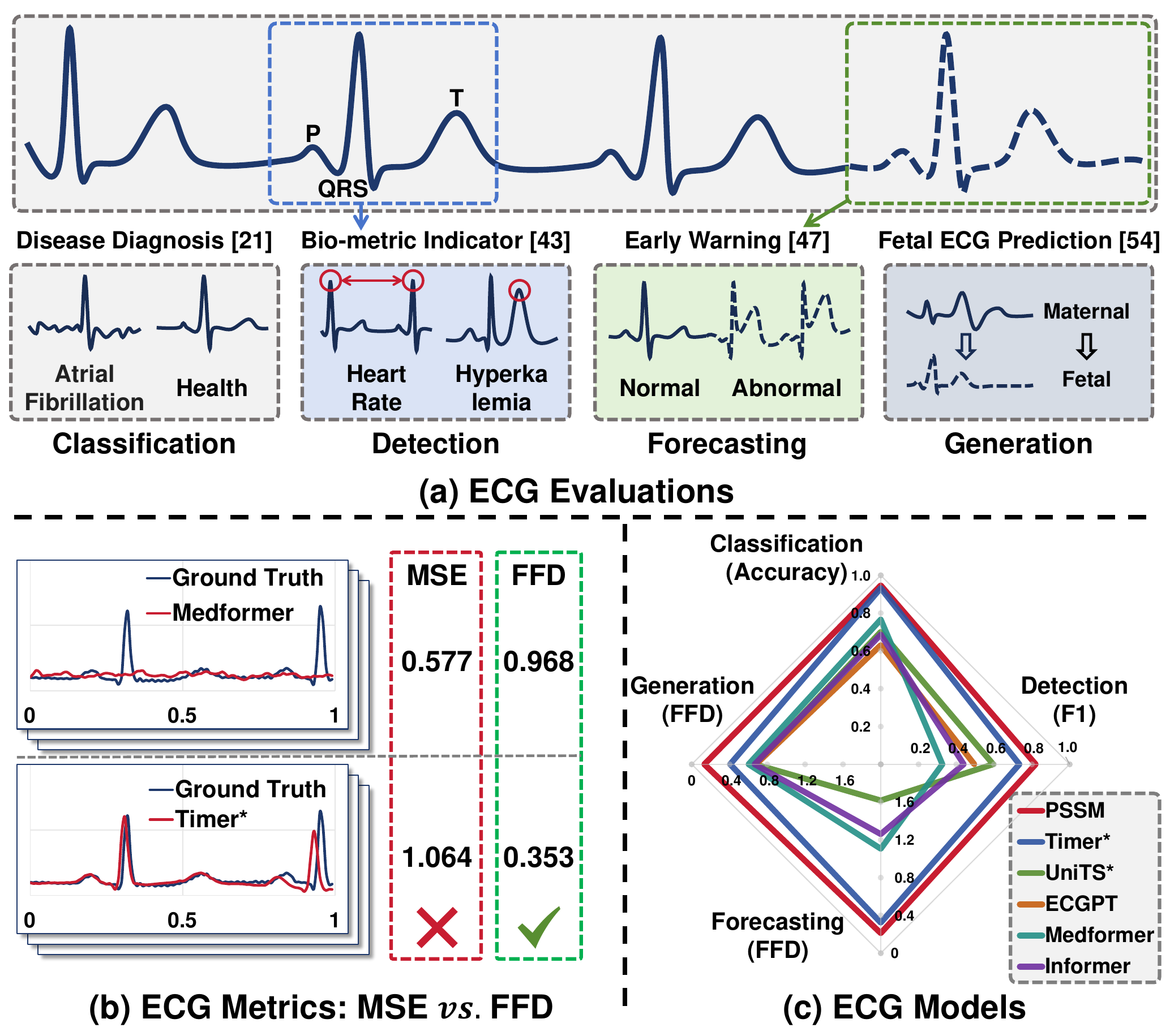}
  \caption{
Illustrations of our proposed ECG benchmark. (a) Overview of four ECG evaluation tasks, including classification, detection, forecasting, and generation, across various ECG applications. (b) Comparison between traditional ECG metric MSE and our proposed metric, FFD. The results are computed using the forecasting results of Medformer~\cite{wang2024medformer} and Timer$^*$~\cite{liutimer} on the NFE dataset~\cite{NFE}; more details are provided in Section~\ref{analysis}. (c) Performance comparison of ECG models, including traditional methods, large time series models, and our proposed Patch Step-by-Step Model~(PSSM).
}
\Description{}
  \label{fig:teaser}
    \vspace{-16pt}
\end{figure}

The electrocardiogram (ECG) is a time-series representation that records the heart's electrical activity, describing the potential variations of myocardial cells during the depolarization and repolarization processes~\cite{bonow2011braunwald}. It is collected through electrodes placed on the chest and limbs~\cite{macfarlane2010comprehensive}. In traditional analysis, the main components of an ECG are the P-wave (atrial depolarization), QRS complex (ventricular depolarization), and T-wave (ventricular repolarization)~\cite{kligfield2007recommendations}, as shown in Figure~\ref{fig:teaser}(a). Abnormalities in these waveforms and their temporal domains can be used to monitor cardiac health, such as arrhythmia classification~\cite{fuster2006acc} and early detection of cardiovascular events~\cite{reichlin2009early}, and to diagnose cardiac diseases, such as myocardial ischemia and electrolyte imbalances~\cite{bonow2011braunwald}. Researchers usually collect ECG data for their research objective, employing traditional methods such as time-frequency analysis~\cite{van1980analysis, goldberger2000physiobank}. However, with the advent of deep learning and large-scale training, the artisanal paradigm has become obsolete due to its limitations, such as insufficient data~\cite{miotto2018deep} and diverse research goals~\cite{rajpurkar2017cardiologist}. Therefore, integrating and standardizing ECG data and its downstream objectives into a unified benchmark has become an urgent requirement for advancing this field.

Inspired by the success of large language models, pre-training large time-series models (LTMs) has become a mainstream trend in time-series modeling~\cite{liutimer,gao2024units}. As a time-series data type, ECG is often integrated with other time-series datasets, such as meteorological and financial data, to pre-train LTMs~\cite{ma2024survey}. However, this integration overlooks the unique characteristics of ECG, leading to potential misunderstandings. Compared to other time-series data, ECG has the following features: (1) It exhibits quasi-periodic behavior, which differs from other time-series with no periodicity (e.g., financial data) or stable periodicity (e.g., weather data). This is because the periodicity of ECG is influenced by physiological factors, such as myocardial ischemia~\cite{wagner2014marriott,chakraborty2018efficient}; (2) ECG is applied in a wide range of complex applications, such as disease diagnosis based on classification and maternal-fetal waveform prediction based on generations. These tasks differ from the primarily predictive goals of other time-series data, like stock price forecasting~\cite{liutimer,zhou2022fedformer}. Therefore, it is necessary to construct a specialized benchmark for ECG to validate existing methods and guide the design of new ones.
In this paper, we propose a comprehensive benchmark to standardize ECG research, comprising the following three parts: 

\textbf{(1) }\textbf{Comprehensive ECG evaluations}. As shown in Figure~\ref{fig:teaser}(a), we define a set of evaluations designed for the broad range of medical applications of ECG, which include four downstream tasks:
(a) Classification for disease diagnosis and event prediction~\cite{hannun2019cardiologist,ribeiro2020automatic},
(b) Detection for key waveform (e.g., P-wave) localization ~\cite{murray1980real,kumar2019efficient},
(c) Forecasting for ECG dynamics prediction~\cite{prakarsha2022time},
(d) Generation for maternal-fetal ECG separation~\cite{sarkar2021cardiogan}.
% These tasks provide a comprehensive framework for evaluating ECG models, ensuring they are rigorously tested and can effectively adapt to various ECG applications.
These tasks provide a comprehensive framework for evaluation, which can fully assess model performance and ensure that they are rigorously tested and can be effectively adapted to various ECG applications.

\textbf{(2) Innovative ECG Metric.}
The Mean Squared Error (MSE), commonly used for time-series evaluations, is often unsuitable for assessing ECG and may lead to misleading results. For example, as shown in Figure~\ref{fig:teaser}(b), generated ECGs that preserve clinically valid morphological patterns but exhibit minor temporal shifts can lead to large MSE, even higher than those of a meaningless flat line prediction. This phenomenon arises from the quasi-periodic nature and extreme values in ECG signals, highlighting the need for more suitable evaluation metrics. Inspired by image quality metrics~\cite{heusel2017gans}, we introduce the Feature-based Fréchet Distance (FFD) to assess the quality of generated ECG signals. FFD calculates the distance between real and generated ECGs by comparing their latent feature distributions, providing a more semantically meaningful measure of generated ECG quality.
Additionally, this approach offers improved robustness by better aligning the evaluation with clinical interpretations of ECG morphology.

\textbf{(3) ECG-Specific Model.} Traditional time-series models may be suboptimal for ECG feature extraction due to its unique characteristics. Inspired by the cardiac conduction system~\cite{trayanova2023computational}, we propose the Patch Step-by-Step Model (PSSM), an encoder-centric hierarchical architecture that adaptively captures cross-scale features from localized waveform segments to global rhythm patterns through iterative signal patching. Extensive experiments demonstrate the effectiveness of our proposed model on the benchmark and metric we have introduced.

Our main contributions can be summarized as follows:
% \vspace{-6pt}
\begin{enumerate}
\item We highlight the unique characteristics of ECG signals compared to other time-series data and analyze the limitations of traditional methods and large time-series models in processing ECG. This underscores the need for a comprehensive ECG benchmark.
\item We propose an ECG benchmark comprising: (a) comprehensive multi-task ECG evaluations, integrating classification, detection, forecasting, and generation to represent the underlying applications of ECG; (b) a novel metric, Feature-based Fréchet Distance (FFD), as a supplement to MSE, to assess semantic fidelity of ECG; (c) a novel architecture, Patch Step-by-Step Model (PSSM), specifically tailored for ECG, which hierarchically encodes ECG through adaptive patching.

\item We conduct extensive experiments to validate the rationale of our ECG benchmark, the robustness of FFD, as well as the effectiveness of our proposed PSSM. The experimental results demonstrate that the proposed benchmark and metric are necessary, and PSSM achieves state-of-the-art performance, providing a solid foundation for advancing this field.
\end{enumerate}
\textbf{Code: }\href{https://github.com/ZhijiangTang/ECG-Benchmark}{https://github.com/ZhijiangTang/ECG-Benchmark}  

% \vspace{-20pt}
\section{Related Works}
\label{sec:relatedworks}

% \subsection{ECG Analysis}
% 2/21 Tang
% 目前, ECG分析可以归类两种方法. 1) 基于信号处理技术的传统方法.  2) 基于深度学习的前沿方法. 

% Tang 2/26
\noindent\textbf{ECG Analysis Methods.} Current ECG analysis can be categorized into two main camps:
(1) Traditional signal processing-based methods. For example, researchers used Fourier analysis and wavelet transform for QRS complex detection~\cite{murray1980real,kumar2019efficient} and applied RR interval variability for arrhythmia classification~\cite{tsipouras2002arrhythmia}.
(2) Deep learning-based methods. For example, they used Tranformer-based frameworks for disease classification~\cite{el2024ecgtransform, miotto2018deep, liu2021deep}, CNN-based frameworks for waveforms detection~\cite{peimankar2021dens}, and GAN-based frameworks for ECG generation~\cite{sarkar2021cardiogan,mohebbian2021fetal}.
However, these methods are designed and validated for specific ECG tasks and cannot fully capture the complex characteristics of ECG signals. In this paper, we propose a comprehensive ECG benchmark that thoroughly evaluates the ability of time-series models to address a wide variety of ECG applications.

\noindent\textbf{Time Series Foundational Models.} 
% % 介绍统计方法ARIMA，介绍深度学习方法RNN，LSTM和Transformer。近年来针对Transformer的改进Informer和Medformer。
Time-series analysis has evolved from classical statistical methods like autoregressive integrated moving average model~\cite{box2015time}, to deep learning approaches such as recurrent neural networks~\cite{rumelhart1986learning}, long short-term memory model~\cite{hochreiter1997long}, and transformers~\cite{vaswani2017attention}, which capture long-range dependencies with recurrent or self-attention mechanisms. Recent advancements have focused on improving transformer-based architectures for time-series tasks: Informer~\cite{zhou2021informer} introduced linear complexity sparse attention mechanisms for long sequences, while Medformer~\cite{wang2024medformer} tailored transformers for medical time-series featuring via multi-scale data fusion.

% 2/21 Tang
% 受大语言模型（LLMs）成功的启发，大规模时序模型（LTMs）已成为时序分析的研究热点。1) 早期探索直接将LLMs应用于时序建模2) 更通用的方法是在大规模时间序列的预训练模型.
\noindent\textbf{Large Time-Series Models (LTMs).} LTMs can be categorized into two main approaches: (1) Direct application of large language models (LLMs) to time-series modelling. For example, OneFitsAll~\cite{zhou2023one} leveraged GPT-2 for time-series forecasting, and Time-LLM~\cite{jin2023time} enhanced zero-shot generalization of LLMs through prompt engineering. 
(2) Pre-training models on large-scale time-series datasets~\cite{dooley2024forecastpfn, ekambaram2023tsmixer, ekambaram2024ttms}. For example, Timer~\cite{liutimer} pre-trained a Transformer decoder using next-token prediction for time-series forecasting, and UniTS~\cite{gao2024units} extended Transformer encoder with multi-task heads. Furthermore, ECGPT~\cite{davies2024interpretable} pre-trained a Transformer decoder on ECG data, while its small pre-training dataset and singular focus limited its ability to address the diverse downstream applications of ECG.
In this paper, we highlight the unique characteristics of ECG compared to other time-series data and argue that LTMs may not necessarily be suitable for various ECG applications. Our extensive experiments demonstrate our benchmark's rationality, our metric's robustness, and our model's effectiveness, offering new insights for the development of ECG analysis.

\begin{figure*}[htbp]
    \centering
    \includegraphics[width=1\linewidth]{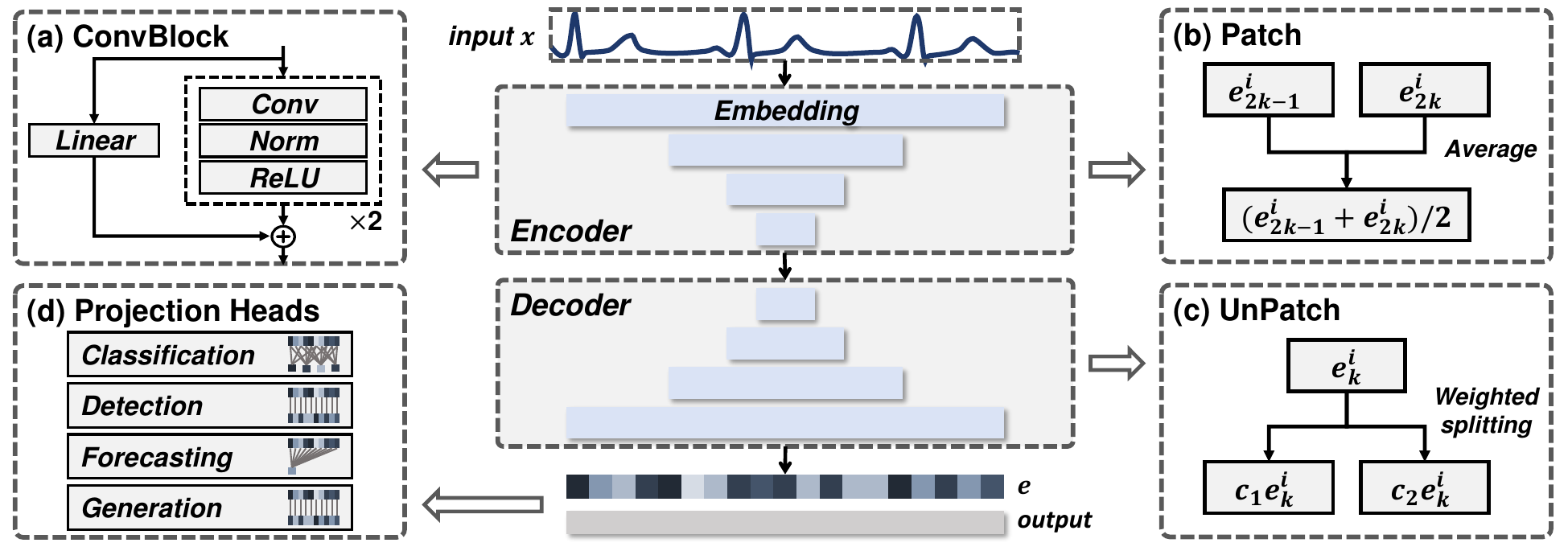}
   \caption{
   Illustrations of the architecture of our Patch Step-by-Step Model (PSSM). (a) Components of the ConvBlock. (b) Patching operation in the encoder, where the tokens are obtained by averaging the corresponding tokens from the previous layer. (c) Unpatching operation in the decoder, where tokens are generated by the weighted splitting of the corresponding tokens from the last layer. (d) Projection heads for the four ECG evaluation tasks.
   }
   \Description{}
    \label{fig:pssm}
    \vspace{-12pt}
\end{figure*}
\section{Method}
\label{sec:method}

\subsection{Preliminary: Time-series Models}
% 时序训练的输入输出，loss，方法
The training of time-series models mimics the training of large language models, leveraging a self-supervised framework based on the next token prediction~\cite{liutimer,gao2024units}. Given a time-series dataset \(\mathcal{D}\!=\!\{\bm{x}\}\), each sample $\bm{x}$ with time length $t$ is initially patched into a sequence of $n$ time tokens, i.e., \(\bm{x}\!=\!\{s_1, s_2, \dots, s_n\}\), where each token spans a time length of $m=t/n$. Then, the loss function for training time-series models can be written as:
\begin{gather}
\label{eq:ntp}
\mathcal{L}_\text{NTP}=\frac{1}{n\!-\!1}\sum_{i=1}^{n-1} \frac{1}{m}\left\| s_{i+1}-e_{i}w \right\|^2,
\end{gather}
where \(s_{i+1} \!\in\! \mathbb{R}^{1 \times m}\) is the ground truth for the \((i\!+\!1)\)-th time-series tokens with \(m\) time length, \(e_i\!\in\! \mathbb{R}^{1 \times d}\) denotes the features of tokens at \(s_i\) extracted by the time-series model, $d$ denotes the hidden dimension, and \(w \in \mathbb{R}^{d \times m}\) is a linear layer aligning the hidden features with the prediction space.

As a type of time-series signal, ECG can also be trained using Eq.~\eqref{eq:ntp}. Given the diverse downstream applications and unique characteristics of ECG, we next introduce our benchmark with four evaluations, a new metric for ECG, and our proposed ECG models.

\subsection{ECG Evaluations}
\label{sec:ECG Evaluations}

\noindent\textbf{Classification.}
ECG can be used to classify diseases, such as atrial fibrillation and hyperkalemia classification, based on distinct ECG patterns from different patients~\cite{macfarlane2010comprehensive}. We thus formulate the classification task as an ECG evaluation, where the input is an individual’s ECG, i.e., $\bm{x}$, and the output is its corresponding class label $y$. The loss can be written as:
\begin{gather}
\label{eq:loss_cls}
\mathcal{L}_\text{cls}= y\log({\frac{\exp(\bar{\bm{e}}w)}{\sum_k\exp(\bar{\bm{e}}w)_k}}),
\end{gather}
where $y$ denotes the one-hot class label, $\exp$ denotes the exponential function, $\bar{\bm{e}} \!\in\! \mathbb{R}^{1 \times d}$ is the averaged feature of $\bm{e}\!=\!\{e_1,e_2,\ldots,e_n\}$, extracted from $\bm{x} \!=\! \{s_1, s_2, \dots, s_n\}$ by models, $d$ is the feature dimension, \(w \!\in\! \mathbb{R}^{d \times c}\) is a linear layer, $c$ is the number of classes, and $k$ indexes all classes.

\noindent\textbf{Detection.}
ECG can be used to calculate cardiovascular metrics (e.g., heart rate variability) by detecting key waveform positions~\cite{peimankar2021dens}. Thus, we formulate detection as an evaluation for ECG, where the given input is $\bm{x}$. The output is the position of key waveforms, annotated with a binary waveform label $\bm{y}$, which spans the same duration $t$ as $\bm{x}$. The loss for detection can be written as:
\begin{gather}
    \label{eq:loss_det}
    \mathcal{L}_{\text{det}} =  \bm{y} \log \left( \sigma(\bm{e}w) \right) + (1 \!-\! \bm{y}) \log \left( 1\!-\!\sigma(\bm{e}w) \right),
\end{gather}
where $\bm{y} \!\in\! \mathbb{R}^{1 \times t}$ is the binary label, $\sigma$ is the sigmoid function, $\bm{e} \!\in\! \mathbb{R}^{n \times d}$ is the extracted feature for $\bm{x}$, and $w \!\in\! \mathbb{R}^{d \times m}$ is the linear projector. Note that, to align with the dimensions of $\bm{y}$, a reshape operation is applied to $\bm{e}w \in \mathbb{R}^{n \times m}$ to convert its dimension into $\mathbb{R}^{nm \times 1}$ (i.e., $\mathbb{R}^{t \times 1}$), which is omitted in the equation for simplicity.

\noindent\textbf{Forecasting.} 
ECG enables early risk alerts and facilitates personalized treatment by predicting its changes~\cite{reichlin2009early}. We thus define the forecasting task, where the input is an ECG time-series \(\bm{x}\!=\!\{s_1, s_2, \dots, s_n\}\), and the output is the subsequent ECG \(\bm{x}'\!=\!\{s_{n+1}, s_{n+2}, \dots, s_{n+n'}\}\), and the loss is similar to Eq.~\eqref{eq:ntp}:
\begin{gather}
\label{eq:forecasting}
\mathcal{L}_\text{forecast}=\frac{1}{n'\!-\!1}\sum_{i=n}^{n+n'\!-\!1} \frac{1}{m}\left\| s_{i+1}-e_{i}w \right\|^2,
\end{gather}
where the notations have the same meanings as in Eq.~\eqref{eq:ntp}.

\noindent\textbf{Generation.}
In clinical practice, acquired ECGs often contain noise \cite{chatterjee2020review} or require invasive acquisition (e.g., fetal ECG \cite{NFE2013}). Therefore, we define the generation task, where the input is an easily accessible ECG $\bm{x}$ with time length $t$, and the output is a corresponding ECG $\bm{y}$ aligned with $\bm{x}$. The loss is defined as:
\begin{gather}
\label{eq:loss_gen}
\mathcal{L}_\text{gen}=\frac{1}{t}\sum_{i=1}^t\left \|y_{i} -e_iw\right \|^2 ,
\end{gather}
where $y_i$ is the ground truth ECG at time $i$ of $\bm{y}$, $e_i \in \mathbb{R}^{1 \times d}$ is the features of $\bm{x}$ at time $i$ extracted by the model, and $w \in \mathbb{R}^{d \times 1}$ is a linear projector.

\subsection{ECG Metric: Feature-based Fréchet Distance}
Mean Squared Error (MSE) is widely applied in temporal forecasting evaluations~\cite{liutimer,gao2024units,zhou2021informer,nie2022time} due to its simplicity and efficiency. However, it exhibits two critical limitations for ECG assessment: (1) it fails to capture the semantic fidelity of ECG (e.g., the quasi-periodic characteristics), and (2) it is sensitive to extreme values (e.g., the R-wave). Figure~\ref{fig:teaser}(b) shows that generated ECG preserving clinically valid patterns yields higher MSE than meaningless flat line predictions when minor temporal shifts occur. Inspired by image quality metrics~\cite{heusel2017gans}, we propose a new metric, Feature-based Fréchet Distance (FFD), as a complementary of MSE.

Consider a mapping $f\!:\!\mathbb{R}^t \!\rightarrow\! \mathbb{R}^k$ that projects ECG signals $\bm{x}$ into normally distributed features $e \sim \mathcal{N}(\mu,\Sigma)$. We quantify the differences between generated and real ECG \textbf{F}eature distributions using the \textbf{F}réchet \textbf{D}istance:
\begin{gather}
\label{eq:ffd}
\scalebox{1}{$\text{FFD}(e, \hat{e}) = \frac{1}{\sqrt{k}} \left( \left\| \mu - \hat{\mu} \right\|^2 + \text{Tr}\left( \Sigma + \hat{\Sigma} - 2 (\Sigma \hat{\Sigma})^{\frac{1}{2}} \right) \right)$},
\end{gather}  
where $\text{Tr}$ is the trace operator, $\mu$ and $\Sigma$ denote the mean vector and covariance matrix of real ECG features, while $\hat{\mu}$ and $\hat{\Sigma}$ denote those of generated ECG features.

To calculate Eq.~\eqref{eq:ffd}, empirically, $\mu,\Sigma$ could be estimated by  $\tilde{\mu},\tilde{\Sigma}$ as follows:
\begin{gather}
\scalebox{1}{$ \tilde{\mu} = \frac{1}{N}\sum_{i=1}^N \bm{e}_i, \quad
 \tilde{\Sigma} = \frac{1}{N-1}\sum_{i=1}^N (\bm{e}_i - \tilde{\mu})^\top(\bm{e}_i - \tilde{\mu})$,
    }
\end{gather}
where $N$ denotes the number of real ECG samples, $\bm{e}_i$ represents the feature of $\bm{x}_i$ mapped by $f$, which could be implemented as a standard time-series feature extractor and the details are given in the experiments. $\hat{\mu}$ and $\hat{\Sigma}$ could be estimated in the same way.

Our mapping $f$ is a pre-trained transformer encoder that uses a mask token prediction object. The loss is as follows:
\begin{gather}
\label{eq:mtp_app}
    \mathcal{L}_{\text{MTP}}=\frac{1}{k}\sum_i\frac{1}{m}\|s_i-e_iw\|^2,
\end{gather}
where $k$ is the number of masked tokens, $s_i$ is the ground truth of $i$-th masked token, \(e_i\!\in\! \mathbb{R}^{1 \times d}\) denotes the features of time $i$ extracted by model, $d$ denotes the hidden dimension, and \(w \in \mathbb{R}^{d \times m}\) is a linear projector.

\subsection{ECG Model: Patch Step-by-Step Model}

\noindent\textbf{Overview.}
As shown in Figure~\ref{fig:pssm}, we introduce the Patch Step-by-Step Model (PSSM) for ECG, applying a hierarchical encoder-decoder architecture. The encoder stacks multiple patch operations and ConvBlock to progressively compress temporal resolution while doubling the hidden dimensions, effectively capturing multi-scale periodic patterns. The decoder reverses the patching process by progressively restoring the temporal resolution and halving the hidden dimensions by learnable unpatching operations.

\noindent\textbf{Patching Encoder.} For the encoder, the ECG signal $\bm{x}$ is first fed into the Embedding layer and then undergoes $l$ hierarchical patching operations, each followed by a ConvBlock. The computation formulations can be written as follows:
% \vspace{-2pt}
\begin{equation}
\label{eq:patch}
\scalebox{1}{$    
\begin{aligned}
        \bm{e}^1&=\text{Embeding}(\bm{x}), \\
        \bm{e}^{i+1}&= \text{ConvBlock}^i(\text{Patch}(\bm{e}^{i})),\\
        % \bm{h}^{i+1}&=\text{ConvBlock}^i(\bm{e}^{i}), 
\end{aligned}$}
\end{equation}
% \vspace{-2pt}
where the Embedding layer maps $\bm{x}\!\in\! \mathbb{R}^{t \times 1}$ to $\bm{e}^1 \!\in\! \mathbb{R}^{t \times d}$, 
% implemented by a Convolutional layer, 
$d$ is the hidden dimension, $i=1,2,\dots,l$ denote the $i$-th layer of the encoder and $l$ denote the number of layers. The $\text{Patch}$ layer splits $\bm{e}^i\in \mathbb{R}^{(t/2^{i-1})\times(2^{i-1}d)}$ into patches $\{(e^i_{2k-1} + e^i_{2k})/2\}$ with the length of $t/2^i$, and then fed to ConvBlock$^i$ to double the hidden dimension and derive $\bm{e}^{i+1}\in \mathbb{R}^{(t/2^{i}) \times (2^{i}d)}$.

\noindent\textbf{Unpatching Decoder.} The encoder output is subsequently fed into the decoder, which performs $l$ unpatching operations before applying a final linear transformation to derive the ECG features. The formulations are:
\begin{equation}
\label{eq:unpatch}
\scalebox{1}{$    
\begin{aligned}
          \bm{e}^{i+1}&= \text{ConvBlock}^{i}({\text{UnPatch}(\bm{e}^{i})}),\\
          \bm{e}&=\bm{e}^{2l+1}w,
\end{aligned}$}
\end{equation}
where $i\!=\!l\!+\!1,\dots,2l$ denotes $l$ decoder layers, $w$ is a linear layer and $\bm{e}$ denotes the final extracted features for ECG. The UnPatch layer restores $\bm{e}^{i+1}$ by weighted copying the features from the last layer and then fed into ConvBlock: $(e^{i+1}_{2k-1}, e^{i+1}_{2k})$$= (\text{ConvBlock}^i(c_1 e^i_{k}), $ $\text{ConvBlock}^i({c_2 e^i_{k}}))$, where $c_1$ and $c_2$ are learnable parameters, and ConvBlock halves the hidden dimension. 

Finally, after a linear projector $w$, the feature $\bm{e}$ of the ECG signal as the output of our PSSM will be further fed into downstream tasks described in Section~\ref{sec:ECG Evaluations}.

\begin{table*}[htbp]
  \centering
    \renewcommand{\arraystretch}{1.1}
    \scalebox{1.1}{
\begin{tabularx}{0.9\textwidth}{l|>{\centering\arraybackslash}X>{\centering\arraybackslash}X>{\centering\arraybackslash}X>{\centering\arraybackslash}X>{\centering\arraybackslash}X>{\centering\arraybackslash}X>{\centering\arraybackslash}X>{\centering\arraybackslash}X}
\toprule
\multicolumn{1}{c|}{\multirow{2}{*}{\textbf{Datasets}}} & {\textbf{Informer}} & \textbf{Medformer} & \textbf{UniTS} & \textbf{UniTS$^*$} & \textbf{ECGPT} & \textbf{Timer} & \textbf{Timer$^*$} & \textbf{PSSM} \\
             &  \cite{zhou2021informer}     & \cite{wang2024medformer}      &  \cite{gao2024units}     &      \cite{gao2024units} &  \cite{davies2024interpretable}     &    \cite{liutimer}   &  \cite{liutimer}     & (Ours)  \\
\hline
\textbf{AF} \cite{AF} & 0.648  & 0.970  & 0.653  & 0.944  & 0.952  & \underline{0.996}  & \textbf{0.999 } & \textbf{0.999 } \\
\textbf{NIFEADB} \cite{NIFEADB} & 0.753  & 0.876  & 0.720  & 0.756  & 0.750  & 0.887  & \underline{0.936}  & \textbf{0.937 } \\
\textbf{CPSC2021} \cite{CPSC2021} & 0.629  & 0.888  & 0.516  & 0.704  & 0.590  & 0.911  &\underline{ 0.965}  & \textbf{0.973 } \\
\textbf{RFAA} \cite{RFAA}& 0.535  & 0.659  & 0.414  & 0.579  & 0.528  & 0.875  & \textbf{0.943 } & \underline{0.937}  \\
\textbf{SPB} \cite{goldberger2000physiobank}& 0.691  & 0.640  & 0.427  & 0.759  & 0.578  & 0.958  & 0\underline{.982}  & \textbf{0.992 } \\
\textbf{CPSC2018} \cite{CPSC2018}& \textbf{0.858 } & 0.563  & 0.277  & 0.463  & 0.383  & 0.630  & 0.768  & \underline{0.843}  \\
\hline

\textbf{Average} & 0.686  & 0.766  & 0.501  & 0.701  & 0.630  & 0.876  & \underline{0.932}  & \textbf{0.947 } \\
\bottomrule
\end{tabularx}%
}
% \vspace{4pt}
\caption{
Test Accuracy on ECG classification datasets. Bold numbers indicate the best performance, underlined numbers indicate the runner-up, and ``*'' denotes that the model is further pre-trained on the ECG datasets.
}
    \vspace{-16pt}
\label{tab:classification}
\end{table*}%

% Table generated by Excel2LaTeX from sheet 'Sheet3'
% \begin{table}[htbp]
%   \centering
%   \caption{Add caption}
%     \begin{tabular}{l|cccccccc}
%     \toprule
%     \multicolumn{1}{|l|}{\textbf{AF}} & 0.6483  & 0.9698  & 0.6531  & 0.9438  & 0.9524  & 0.9960  & \textbf{0.9988 } & \textbf{0.9988 } \\
%     \multicolumn{1}{|l|}{\textbf{NIFEADB}} & 0.7535  & 0.8758  & 0.7196  & 0.7558  & 0.7498  & 0.8869  & 0.9360  & \textbf{0.9368 } \\
%     \multicolumn{1}{|l|}{\textbf{CPSC2021}} & 0.6295  & 0.8882  & 0.5161  & 0.7041  & 0.5902  & 0.9108  & 0.9650  & \textbf{0.9730 } \\
%     \multicolumn{1}{|l|}{\textbf{RFAA}} & 0.5355  & 0.6589  & 0.4140  & 0.5789  & 0.5283  & 0.8755  & \textbf{0.9435 } & 0.9372  \\
%     \multicolumn{1}{|l|}{\textbf{SPB}} & 0.6915  & 0.6398  & 0.4265  & 0.7588  & 0.5784  & 0.9579  & 0.9819  & \textbf{0.9920 } \\
%     \multicolumn{1}{|l|}{\textbf{CPSC2018}} & \textbf{0.8584 } & 0.5633  & 0.2766  & 0.4634  & 0.3831  & 0.6304  & 0.7679  & 0.8431  \\
%     \midrule
%     \textbf{Average} & 0.6861  & 0.7660  & 0.5010  & 0.7008  & 0.6304  & 0.8762  & 0.9322  & \textbf{0.9468 } \\
%     \bottomrule
%     \end{tabular}%
%   \label{tab:addlabel}%
% \end{table}%

\begin{table*}[htbp]
  \centering
    \renewcommand{\arraystretch}{1.1}
    \scalebox{1.1}{
\begin{tabularx}{0.9\textwidth}{l|>{\centering\arraybackslash}X>{\centering\arraybackslash}X>{\centering\arraybackslash}X>{\centering\arraybackslash}X>{\centering\arraybackslash}X>{\centering\arraybackslash}X>{\centering\arraybackslash}X>{\centering\arraybackslash}X>{\centering\arraybackslash}X}
\toprule
\multicolumn{1}{c|}{\multirow{2}{*}{\textbf{Dataset}}} & \textbf{GQRS} & \textbf{Informer} & \textbf{Medformer} & \textbf{UniTS} & \textbf{UniTS$^*$} & \textbf{ECGPT} & \textbf{Timer} & \textbf{Timer$^*$} & \textbf{PSSM} \\
     &   \cite{GQRS}   &  \cite{zhou2021informer}     & \cite{wang2024medformer}      &  \cite{gao2024units}     &      \cite{gao2024units} &  \cite{davies2024interpretable}     &    \cite{liutimer}   &  \cite{liutimer}     & (Ours) \\
\hline
\textbf{MITDB} \cite{MITDB}& 0.358  & 0.707  & 0.181  & 0.740  & 0.810  & 0.760  & 0.843  & \underline{0.919}  & \textbf{0.921 } \\
\textbf{SVDB} \cite{SVDB}& 0.527  & 0.539  & 0.746  & 0.792  & 0.859  & 0.728  & 0.900  & \underline{0.939}  & \textbf{0.953 } \\
\textbf{NFE} \cite{NFE}& 0.139  & 0.269  & 0.222  & 0.290  & 0.274  & 0.397  & 0.339  & \underline{0.423}  & \textbf{0.664 } \\
\textbf{FEPL} \cite{FEPL}& 0.246  & \underline{0.848}  & 0.577  & 0.730  & 0.701  & 0.672  & 0.658  & {0.803}  & \textbf{0.978 } \\
\textbf{CPSC2020} \cite{CPSC2020}& 0.049  & 0.153  & 0.117  & 0.147  & 0.308  & 0.050  & 0.421  & \textbf{0.528 } & \underline{0.505}  \\
\textbf{MITPDB} \cite{MITPDB}& 0.090  & 0.119  & 0.115  & 0.346  & 0.619  & 0.364  & 0.667  & \underline{0.791}  & \textbf{0.901 } \\
\hline
\textbf{Average} & 0.235  & 0.439  & 0.326  & 0.508  & 0.595  & 0.495  & 0.638  &\underline{ 0.734 } & \textbf{0.820 } \\
\bottomrule
\end{tabularx}%
}
% \vspace{4pt}
  \caption{
  Test F1 score on ECG detection datasets. Bold numbers indicate the best performance, underlined numbers indicate the runner-up, and ``*'' denotes that the model is further pre-trained on the ECG datasets. GQRS~\cite{GQRS} is a traditional ECG method specifically designed for ECG detection, thus its results are reported only in this table.
  }
      \vspace{-16pt}
  \label{tab:detection}%
\end{table*}%

\section{Experiments}
\label{sec:experiments}

In this section, we first introduce the datasets for each task, then explain the implementation details, and finally analyze the proposed benchmark and method in detail.
\subsection{Dataset}
Building upon prior works \cite{davies2024interpretable,sarkar2021cardiogan}, we collected ECG evaluation datasets from: The China Physiological Signal Challenge 2018-2021 (CPSC2018, CPSC2019, CPSC2020, CPSC2021)~\cite{CPSC2018, CPSC2019, CPSC2020, CPSC2021}, PhysioNet~\cite{goldberger2000physiobank}, as well as other ECG research datasets, with detailed descriptions following.

\noindent\textbf{Classification.} 
We collected 6 datasets for classification:
\begin{itemize}
    \item CPSC2018 contains 7 disease classes. 
    \item CPSC2021 is recorded from 12-lead Holter or 3-lead wearable ECG monitoring devices. Challenge focuses on atrial fibrillation diagnosis and has 3 classes in total.
    \item AF Classification (AF)~\cite{AF} contains short-term ECG recordings with binary class for atrial fibrillation.
    \item Non-Invasive Fetal ECG Arrhythmia (NIFEADB)~\cite{NIFEADB} uses fetal ECGs for binary classification of arrhythmias.
    \item Reducing False Arrhythmia Alarms (RFAA)~\cite{RFAA} designed to classify arrhythmia alarms in ICUs into 4 classes.
    \item St. Petersburg Arrhythmia Database~(SPB)~\cite{goldberger2000physiobank} comprises 75 ECG recordings annotated with 6 diseases.
\end{itemize}

\noindent\textbf{Detection.}
We evaluate 6 datasets for detection:

\begin{itemize} 
    \item Fetal ECG with Pregnancy and Labor~(FEPL)~\cite{FEPL} collects fetal ECGs and maternal ECGs annotated with fetal R-waves.
    \item CPSC2020 was collected from arrhythmia patients with abnormal T-wave annotations.
    \item MIT-BIH Arrhythmia Database~({MITDB}, {MITPDB})~\cite{MITDB,MITPDB}, MITDB contains 48 ECG records, and MITPDB complements MITDB by adding P waves annotations.
    \item Noninvasive Fetal ECG~(NFE)~\cite{NFE} collects maternal abdominal ECGs annotated with fetal R-waves.
    \item Supraventricular Arrhythmia Database (SVDB)~\cite{SVDB} contains 78 ECG records, annotated with QRS complexes.
\end{itemize}
\begin{table*}[ht]
\centering
   \renewcommand{\arraystretch}{1.1}
    \scalebox{1.1}{
\begin{tabularx}{0.9\textwidth}{l|>{\centering\arraybackslash}X>{\centering\arraybackslash}X>{\centering\arraybackslash}X>{\centering\arraybackslash}X>{\centering\arraybackslash}X>{\centering\arraybackslash}X>{\centering\arraybackslash}X}
\toprule
\multicolumn{1}{c|}{\multirow{2}{*}{\textbf{Datasets}}} & \textbf{Informer} & \textbf{Medformer} & \textbf{UniTS} & \textbf{UniTS$^*$} & \textbf{Timer} & \textbf{Timer$^*$} & \textbf{PSSM} \\
      &  \cite{zhou2021informer}     & \cite{wang2024medformer}      &  \cite{gao2024units}     &      \cite{gao2024units}       &    \cite{liutimer}   &  \cite{liutimer}     & (Ours) \\
\hline
\textbf{CPSC2019} \cite{CPSC2019}& 2.241 & 1.866 & 4.172 & 1.975 & 1.988 & \underline{0.839} & \textbf{0.480} \\
\textbf{DALIA} \cite{DALIA}& 1.419 & 1.629 & 4.838 & 2.184 & 1.938 & \underline{0.301} & \textbf{0.182} \\
\textbf{RDBH} \cite{RDBH}& 1.344 & 1.386 & 3.651 & 1.690 & 1.056 & \underline{0.275} & \textbf{0.153} \\
\textbf{MIMICSub} \cite{MIMIC3Sub}& 0.538 & 0.384 & 4.306 & 2.093 & 1.091 & \textbf{0.100} & \underline{0.117} \\
\textbf{NFE} \cite{NFE}& 1.400 & 1.064 & 2.196 & 1.089 & 1.289 & \underline{0.353} & \textbf{0.311} \\
\textbf{PTB} \cite{PTB}& 0.618 & 0.301 & 1.491 & 0.667 & 0.480 & \underline{0.065} & \textbf{0.025} \\
\hline
\textbf{Average} & 1.260 & 1.105 & 3.442 & 1.616 & 1.307 & \underline{0.322} & \textbf{0.211} \\
\bottomrule
\end{tabularx}%

}
% \vspace{4pt}
\caption{
Test FFD on ECG forecasting datasets. A smaller FFD is better. Bold numbers indicate the best performance, underlined numbers indicate the runner-up, and ``*'' indicates that the model is further pre-trained on the ECG datasets. ECGPT~\cite{davies2024interpretable} is not applicable for this setting, and its performance is omitted.
}
    \vspace{-16pt}
\label{tab:forecast}
\end{table*}

\begin{table*}[htbp]
  \centering
        \renewcommand{\arraystretch}{1.1}
    \scalebox{1.1}{
\begin{tabularx}{0.9\textwidth}{l|>{\centering\arraybackslash}X>{\centering\arraybackslash}X>{\centering\arraybackslash}X>{\centering\arraybackslash}X>{\centering\arraybackslash}X>{\centering\arraybackslash}X>{\centering\arraybackslash}X>{\centering\arraybackslash}X}
\toprule
\multicolumn{1}{c|}{\multirow{2}{*}{\textbf{Datasets}}} & \textbf{Informer} & \textbf{Medformer} & \textbf{UniTS} & \textbf{UniTS$^*$} & \textbf{ECGPT} & \textbf{Timer} & \textbf{Timer$^*$} & \textbf{PSSM} \\
      &  \cite{zhou2021informer}     & \cite{wang2024medformer}      &  \cite{gao2024units}     &      \cite{gao2024units} &  \cite{davies2024interpretable}     &    \cite{liutimer}   &  \cite{liutimer}     & (Ours)  \\
\hline
\textbf{MITDB} \cite{MITDB,moody1984noise}& 0.105  & 0.174  & 0.151  & 0.162  & 0.082  & 0.123  & \underline{0.068}  & \textbf{0.016 } \\
\textbf{PTBXL} \cite{PTBXL,moody1984noise}& 0.077  & 0.176  & 0.124  & 0.138  & 0.095  & 0.133  & \underline{0.054}  & \textbf{0.019 } \\
\textbf{ADFECGDB} \cite{ADFECGDB}& 0.495  & 0.347  & 0.430  & 0.417  & 0.457  & 0.323  & \underline{0.321}  & \textbf{0.249 } \\
\textbf{FEPL}  \cite{FEPL} & 0.982  & 1.071  & 0.911  & 1.304  & 1.219  & 1.344  & \underline{0.511}  & \textbf{0.081 } \\
\textbf{BIDMC} \cite{BIDMC}& 1.562  & 1.049  & 1.033  & 1.275  & 1.506  & 1.026  & \underline{0.959}  & \textbf{0.218 } \\
\textbf{SST} \cite{SST}& 0.772  & 0.793  & 0.747  & 0.731  & 0.728  & 0.699  & \underline{0.526}  & \textbf{0.216 } \\
\hline
\textbf{Average} & 0.665  & 0.601  & 0.566  & 0.671  & 0.681  & 0.608  & \underline{0.407}  & \textbf{0.133 } \\
\bottomrule
\end{tabularx}%
}
% \vspace{4pt}
\caption{
Test FFD on ECG generation datasets. A smaller FFD is better. Bold numbers indicate the best performance, underlined numbers indicate the runner-up, and ``*'' indicates that the model is further pre-trained on the ECG datasets.
}
    \vspace{-16pt}
  \label{tab:generation}%
  % \vspace{-4pt}
\end{table*}%

\noindent\textbf{Forecasting.}
While any ECG datasets can be used for forecasting, for broader validation, we selected 6 datasets designed for different research objectives:
\begin{itemize}
    \item CPSC2019~\cite{CPSC2019} includes 2,000 single-lead ECG recordings collected from patients with cardiovascular disease. It is often used in ECG detection tasks.
    \item PPG-DaLiA (DALIA)~\cite{DALIA} contains 
    % physiological and motion data recorded from wrist- and chest-worn devices from 15 subjects while performing various activities. 
    Contains ECG, photoplethysmography, and 3D accelerometer data. It is often used in ECG generation tasks.
    \item PTB Diagnostic ECG Database (PTB, PTBXL)~\cite{PTB,PTBXL}, PTB contains 549 records from 290 subjects. PTBXL ECG contains 21,799 clinical 12-lead ECGs from 18,869 patients. They are often used in ECG classification tasks.
    \item A subset of MIMIC-III Waveform Database Matched Subset~(MIMICSub)~\cite{MIMIC3Sub}, the original dataset contains 22,247 numerical records of 10,282 different ICU patients. We selected a portion of this dataset.
    % We selected the first ECG recording of each patient in the folders ``p00'' and ``p03''.
    \item Detection of Heart Beats Dataset~(RDBH)~\cite{RDBH} contains 10-minute recordings. Each recording contains four to eight signals. 
    It is often used in ECG detection tasks.
    \item European ST-T Database~(EBD)~\cite{EBD} contains 90 annotated Holter recordings from 79 subjects.
\end{itemize}

\noindent\textbf{Generation.}
Generation evaluation contains 6 datasets, including FEPL. Like previous work\cite{chatterjee2020review}, we augment the MITDB and PTBXL\cite{PTBXL} by injecting noise from the MIT-BIH Noise Stress Test Database \cite{moody1984noise} to build denoising datasets. The remaining datasets are described below:

\begin{itemize}
    \item Abdominal and Direct Fetal ECG (ADFECGDB)~\cite{ADFECGDB} contains maternal abdominal ECG and direct fetal ECG.
    \item BIDMC PPG and Respiration Dataset (BIDMC)~\cite{BIDMC} extracts ECG from the MIMIC II Waveform Database~\cite{saeed2011multiparameter}, with annotated photoplethysmogram signals.
    \item SensSmartTech Database (SST)~\cite{SST} includes synchronized cardiovascular signals from ECG, phonocardiograms, and photoplethysmography.
\end{itemize}

We further organized an extension ECG Dataset for LTMs pre-training, comprising AF, CPSC2018, SPB, PTB, PTBXL, NFE, MIMICSub, The Georgia 12-lead ECG Challenge Database~\cite{goldberger2000physiobank}, and the Shaoxing and Ningbo Hospital ECG Database~\cite{Shaoxing, Ningbo}. This dataset contains 98,513 subjects processed into 8,414,834 training samples. 
Although the extension ECG Dataset overlaps with other datasets, the training objectives differ (e.g., AF and CPSC2018), or the overlapping samples are removed (e.g., MIMICSub).
We describe more details about ECG datasets in the supplementary material.

\begin{figure*}[htbp]
    \centering
    \includegraphics[width=1\linewidth]{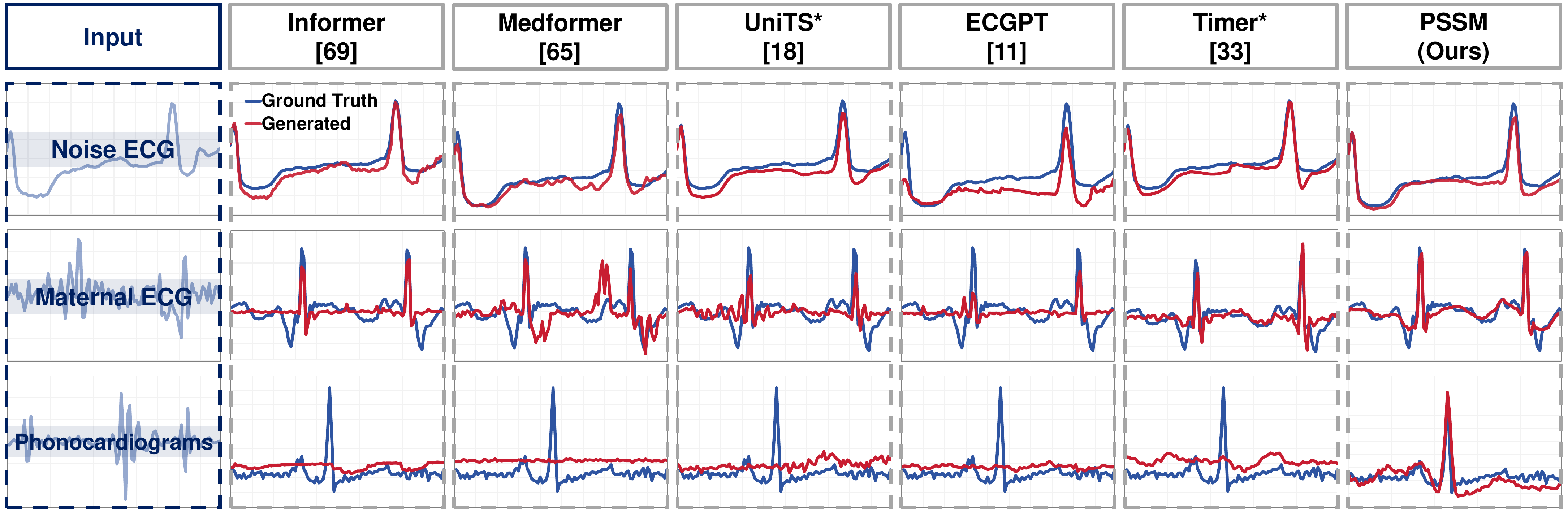}
    \caption{Qualitative results of different models on the ECG generation task. The three rows of samples were selected from the MITDB~\cite{MITDB}, FEPL~\cite{FEPL}, and SST~\cite{SST} datasets, respectively. Red curves are the ground truth ECG, and blue curves are the generated ECG.
    }
        \vspace{-16pt}
    \Description{}
    \label{fig:generation_plot}
\end{figure*}

\subsection{Implementation Details}
\noindent\textbf{Datasets.}
All ECG recordings were resampled to 100Hz with 500 points fixed length. We discarded recordings with $>25\%$ missing data or abnormal morphologies, imputing missing segments via linear interpolation. Multi-channel ECG recordings were split into single-channel recordings to standardize input dimensions. The datasets were partitioned into 50\% for training/fine-tuning and 50\% for testing. All datasets have been stripped of sensitive information.

We addressed class imbalance by removing underrepresented classes and downsampling overrepresented ones for classification. In detection tasks with severe class imbalance (waveform annotation positions <5\%), we adopted the F1-score. 
Non-maximum suppression was applied to eliminate redundant waveform probabilities. The predicted position must be within a $\pm$70ms window centred on the annotation position\cite{CPSC2020,association1998testing}. For forecasting, the predicting length is 100.

\noindent\textbf{Models.}
We evaluated advanced LTMs introduced in Section~\ref{sec:relatedworks}: Timer~\cite{liutimer}, UniTS~\cite{gao2024units}, and ECGPT~\cite{davies2024interpretable}. Both raw versions and variants are pre-trained on the extension ECG Dataset, denoted by ``*''. The mapping $f$ of FFD is the LTM constructed by the transformer encoder. Additionally, we tested state-of-the-art transformer-based end-to-end models: Informer~\cite{zhou2021informer} and Medformer~\cite{wang2024medformer} alongside our proposed PSSM.
All pre-trainings use the extension ECG Dataset and were conducted on NVIDIA RTX 4090 GPUs with a batch size of 8192 over 300 epochs. We use the Adam optimizer~\cite{kingma2014adam} for optimization with a learning rate of $1 \times 10^{-6}$.
All test experiments were conducted on NVIDIA A800 GPUs with a batch size of 1024 over 100 epochs, using the Adam optimizer with learning rates empirically selected from $\{1\!\times\!10^{-4}, 5\!\times\!10^{-5}, 1\!\times\!10^{-5}\}$. For LTMs, we freeze the backbone and only fine-tune a fully connected layer and task heads, yet remain unmodified in forecasting. Other models undergo full training. More details are provided in the supplementary material.

\subsection{Analysis}
\label{analysis}
Through the following Q\&A, we provide an in-depth analysis of the experimental results, focusing on the rationale of our ECG benchmark, the robustness of our proposed FFD, and the effectiveness of our PSSM.

\vspace{6pt}
\noindent\textbf{Q1.} \textit{\textbf{Is the proposed ECG benchmark reasonable and is the PSSM effective?}}\\
\noindent\textbf{A1.} 
The main results for our four evaluation tasks, i.e., classification, detection, forecasting, and generation, are presented in Table~\ref{tab:classification} to Table~\ref{tab:generation}, respectively. Our proposed method, PSSM, achieves state-of-the-art performance across all tasks, with an average performance of 0.947 in classification accuracy, 0.820 in detection F1 score, 0.211 in forecasting FFD, and 0.133 in generation FFD. 

Compared to traditional methods such as Medformer, PSSM achieves an averaged relative improvement of 83.4\%, specifically 23.6\% in classification accuracy, 151.2\% in detection F1 score, 80.9\% in forecasting FFD, and 80.9\% in generation FFD. 
Compared to large time-series methods such as Timer, PSSM achieves an averaged relative improvement of 49.6\%, specifically 8.1\% in classification accuracy, 28.6\% in detection F1 score, 83.8\% in forecasting FFD, and 78.1\% in generation FFD.
Compared to rule-based GQRS in detection, PSSM improves significantly by 151.2\%. 
These results demonstrate the effectiveness of our method and highlight the need to develop specialized models for ECG analysis rather than relying solely on large time-series models.

These experimental results further validate the robustness and consistency of our proposed ECG benchmark. For example, our method PSSM consistently achieves the highest average performance, while Timer$^*$ consistently secures the runner-up position, demonstrating the consistency of four evaluations. Even when some models perform well on specific datasets (e.g., Informer on CPSC2018 classification), our comprehensive benchmark effectively mitigates these outlier cases, demonstrating robustness.

% \vspace{6pt}
\noindent\textbf{Q2.} \textit{\textbf{Can FFD effectively assess ECG quality?}}\\
\noindent\textbf{A2.} 
Our proposed FFD is a reliable metric for assessing the quality of generated ECGs, with lower FFD values indicating better ECG fidelity. As illustrated in the third row of Figure~\ref{fig:generation_plot}, PSSM accurately reconstructs the R-wave location and amplitudes from phonocardiograms on the SST dataset, whereas other models yield meaningless flat lines. This finding aligns with the quantitative results in Table~\ref{tab:generation}, where PSSM achieves the best FFD as 0.216, significantly outperforming other methods, such as Timer$^*$ by 0.31 (59\%). This consistency between qualitative and quantitative results is also validated in the first-row MITDB and second-row FEPL datasets. 

\begin{figure}
    \centering
    \includegraphics[width=\linewidth]{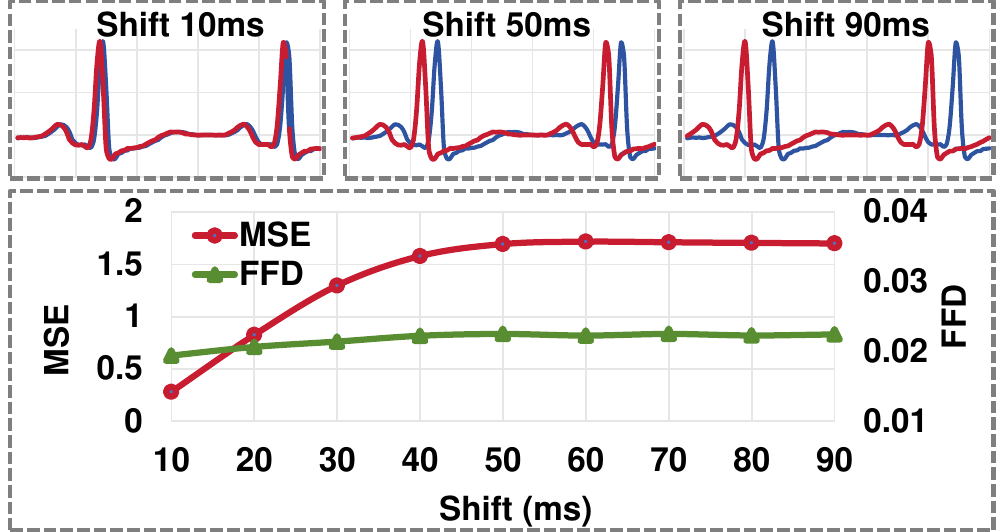}
    \caption{
    Comparisons for MSE and FFD under different temporal shifts. The top row displays ECG corresponding to various temporal shifts, where red curves are the ground truth and blue curves are the ECG with time shift. The bottom row illustrates how MSE and FFD vary across different temporal shifts. As temporal shift increases, although the semantic meaning of ECG remains consistent, the MSE increases while the FFD remains stable.    }
    \Description{}
    \vspace{-16pt}
    \label{fig:shift_mse_vs_ffd}
\end{figure}

Additionally, FFD shows robustness in evaluating the semantics of generated ECGs. As illustrated in Figure~\ref{fig:shift_mse_vs_ffd}, ECGs preserve diagnostic semantics under temporal shifts in the top row, with FFD remaining stable at 0.022 despite perturbations, as shown in the bottom. In contrast, the MSE sharply increases from 0.281 to 1.579 under the same shifts, failing to capture the unchanged semantics of ECG signals. This demonstrates that FFD is robust to temporal fluctuations, only related to ECG semantics.

In conclusion, FFD is an effective and robust metric for assessing the quality of ECGs. For completeness, in the supplementary materials, we supplemented the MSE performance in the forecasting and generation tasks, where our PSSM is still the best model, and we also provided more detailed discussions on MSE and FFD.

% \vspace{6pt}
\label{QA:q3}
\noindent\textbf{Q.3} \textit{\textbf{How does hierarchical patching improve PSSM?}}\\
\noindent\textbf{A.3} 
To validate the hierarchical patching strategy, as shown in Table~\ref{tab:ablation}, we conducted two ablation studies: (1) replacing ConvBlocks with a transformer encoder (``rp trans'') and (2) applying patching only at the input stage, similar to LTMs, instead of using hierarchical patching (``w/o ssp''). The results indicate that PSSM with ConvBlocks and hierarchical patching achieves the best performance, outperforming the ``rp trans'' variant by 63.2\%, 159.8\%, 85.6\%, and 87.9\% across the four evaluation tasks, respectively, and outperforming the ``w/o ssp'' variant by 9.6\%, 70.06\%, 78.3\%, and 62.4\% across the four evaluation tasks. These results demonstrate that the hierarchical patching strategy is both necessary and effective. The possible reason for the failure of the transformer encoder is that hierarchical patching could capture the quasi-periodic temporal patterns for accurate feature extraction, while ``rp trans'' variants create complex feature interactions by self-attention along the time dimension that are redundant for periodic signals.

% \vspace{6pt}
\noindent\textbf{Q.4} \textit{\textbf{Are LTMs suitable for ECG tasks?}}\\
\noindent\textbf{A.4} 
Based on our experiments, the performance of LTMs is less than ideal. As shown in Table~\ref{tab:classification} to Table~\ref{tab:generation} and Figure~\ref{fig:generation_plot}, LTMs, including Timer, Timer$^*$, UniTS, UniTS$^*$, and ECGPT, exhibit suboptimal performance in our ECG benchmark. Specifically, (a) the raw LTMs (i.e., Timer and UniTS) significantly underperform (e.g., UniTS and Timer show accuracy drops of 0.446 and 0.071 compared to PSSM in classification). (b) ECGPT, an LTM specifically pre-trained on ECG datasets, also struggles (e.g., with a 39.6\% lower F1-score than PSSM in detection). (c) The LTMs further pre-trained on ECG datasets, e.g., Timer$^*$ outperform Timer by 75.3\% in forecasting FFD yield 34.5\% lower than PSSM. Demonstrating some improvement compared to their raw versions, but still do not match the performance of our proposed model.

\begin{figure}[tp]
    \centering
    \includegraphics[width=\linewidth]{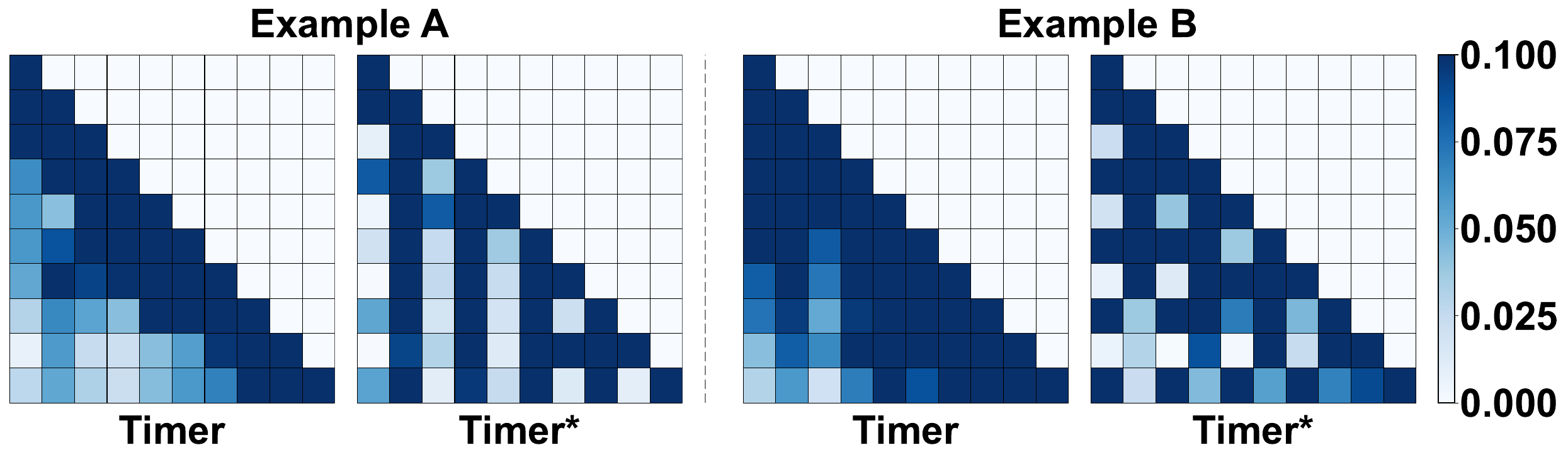}
    \caption{
    Illustrations of the attention maps for a pre-trained time-series model with and without additional ECG pre-training. Timer$^*$, which is further pre-trained on ECG datasets, exhibits an attention map with periodic patterns not observed in the raw Timer model. The samples are selected from the PTB dataset~\cite{PTB}.
    }
        \vspace{-16pt}
    \Description{}
    \label{fig:attention_map}
\end{figure}

On the other hand, the qualitative results show that raw LTMs fail to capture the periodic nature of ECG signals. Figure~\ref{fig:attention_map} reveals that the raw Timer model attends only to nearby tokens, whereas Timer$^*$ (further pre-trained on ECG data) adjusts its attention to capture periodic patterns. For example, Timer$^*$ focuses on even tokens in Example A and alternates between even and odd tokens in Example B. This phenomenon indicates that the underlying attention mechanism of transformers in current LTMs is unsuitable for ECG data. Even when further pre-trained on ECG datasets to learn more ECG patterns like Timer$^*$, transformers still fail to fully leverage their potential (i.e., half of the attention weights are wasted), resulting in suboptimal performance. Therefore, an improved model architecture is necessary to achieve better performance for ECG analysis.
% 在补充材料中，我们也进一步提供了LTMs不适合ECG任务的证据和讨论。
In the Supplementary Material, we also provide further evidence and discussion of the unsuitability of LTMs for ECG tasks.

\begin{table}[tp]
    \centering
        \renewcommand{\arraystretch}{1.1}
    \scalebox{1.1}{
\begin{tabularx}{0.9\linewidth}{l|>{\centering\arraybackslash}X>{\centering\arraybackslash}X>{\centering\arraybackslash}X>{\centering\arraybackslash}X}
\toprule
\multirow{2}{*}{\textbf{Model}}& \textbf{CLS} & \textbf{DET} & \textbf{FCAST} & \textbf{GEN} \\
& {(Accuracy)} & {(F1)} & {(FFD)} & {(FFD)} \\
\hline
\textbf{\quad rp trans} & 0.582  & 0.316  & 1.198  & 1.101  \\
\textbf{\quad w/o ssp} & 0.866  & 0.483  & 0.791  & 0.353  \\
\hline
\textbf{PSSM} & \textbf{0.949 } & \textbf{0.820 } & \textbf{0.172 } & \textbf{0.133 } \\
\bottomrule
\end{tabularx}%
    }
    \caption{
    Ablation studies of our PSSM on four ECG evaluations, where CLS, DET, FCAST, and GEN denote classification, detection, forecasting, and generation, respectively. ``rp trans'' substitutes ConvBlocks with Transformer encoders, and ``w/o ssp'' has no hierarchical patch.
    }
    \vspace{-16pt}
    \label{tab:ablation}
\end{table}

\section{Conclusion}
\label{sec:conclusion}
This paper introduces a comprehensive benchmark for ECG analysis, including classification, detection, forecasting, and generation, designed to address real-world clinical needs for ECG, such as disease diagnosis and early risk alerts. 
Second, we propose a novel metric for ECG, Feature-based Fréchet Distance (FFD), which addresses the limitations of Mean Squared Error (MSE) in ECG assessment, particularly in capturing the semantic fidelity of ECG.
Additionally, we introduce the Patch Step-by-Step Model (PSSM), an ECG-specialized architecture that employs a hierarchical patching strategy to effectively capture the nature of ECG signals. Extensive experiments validate the rationality of our ECG benchmark, the robustness of our proposed FFD, and the superiority of our PSSM compared to other state-of-the-art time-series methods across all evaluation tasks. Our future work focuses on developing ECG-optimized time-series model pre-training to bridge the gap between general LTM pretraining and ECG requirements.

\section{Acknowledgments}
This work was supported by the Strategic Priority Research Program of the Chinese Academy of Sciences under Grant No. XDA0460205.

% \section{Problem statement}

% In this document we discuss how to write an ACM article.

% \section{Methods}

% This document provides \LaTeX\ templates for the article. We
% demonstrate different versions of ACM styles and show various options
% and commands.  We add extensive documentation for these commands and
% show examples of their use.

% \section{Results}

% We hope the resulting templates and documentation will help the
% readers to write submissions for ACM journals and proceedings.

% \section{Significance}

% This document is important for anybody wanting to comply with the
% requirements of ACM publishing.
% \bibliographystyle{ACM-Reference-Format}
% \bibliography{ref}

\bibliographystyle{ACM-Reference-Format}
\balance
\bibliography{ref}

\newpage
\section*{Supplementary Material}
The supplementary material provides additional details to complement the main paper. \textbf{More details for method}, we elaborate on the complete outputs of classification and detection, the approach to constructing the generation denoise dataset, and the FFD mapping $f$ training method. \textbf{More details for datasets}, we offer more details about their characteristics and scales. \textbf{More details for implementation details}, we supplement the parameter settings of each model. \textbf{More details for results and analysis}, we present the MSE results of forecasting and generation, along with a more comprehensive analysis.
\subsection*{Method}
% \subsubsection*{ECG Evaluations}

\noindent\textbf{Classification}.
For the ECG classification task, we re-write the loss as follows:
\begin{gather}
\label{eq:loss_cls_app}
\mathcal{L}_\text{cls}= y\log({\frac{\exp(\bar{\bm{e}}w)}{\sum_k\exp(\bar{\bm{e}}w)_k}}), \tag{a}
\end{gather}
where $y$ denotes the one-hot class label, $\exp$ denotes the exponential function, $\bar{\bm{e}} \!\in\! \mathbb{R}^{1 \times d}$ is the averaged feature of $\bm{e}\!=\!\{e_1,e_2,\ldots,e_n\}$, extracted from $\bm{x} \!=\! \{s_1, s_2, \dots, s_n\}$ by models, $d$ is the feature dimension, \(w \!\in\! \mathbb{R}^{d \times c}\) is a linear layer, $c$ is the number of classes, and $k$ indexes all classes.

While Eq.~\eqref{eq:loss_cls_app} is a training loss, the formula for obtaining the predicted category $\hat{y}$ is as follows:
\begin{gather}
    \hat{y} = \underset{i}{\operatorname{argmax}}\left(\frac{\exp(\bar{\bm{e}}w)_i}{\sum_k\exp(\bar{\bm{e}}w)_k}\right), \tag{b}
\end{gather}
where $\hat{y}$ denotes the predicted class label. The $\operatorname{argmax}$ operator selects the class index $i$ with the highest probability $\exp(\bar{\bm{e}} w)_i$ among all candidate classes.

\noindent\textbf{Detection}.
For the ECG detection task, we re-write the loss as follows:
\begin{gather}
    \label{eq:loss_det_app}
    \mathcal{L}_{\text{det}} =  \bm{y} \log \left( \sigma(\bm{e}w) \right) + (1 \!-\! \bm{y}) \log \left( 1\!-\!\sigma(\bm{e}w) \right), \tag{c}
\end{gather}
where $\bm{y} \!\in\! \mathbb{R}^{1 \times t}$ is the binary label, $\sigma$ is the sigmoid function, $\bm{e} \!\in\! \mathbb{R}^{n \times d}$ is the extracted feature for $\bm{x}$, and $w \!\in\! \mathbb{R}^{d \times m}$ is the linear projector. Note that, to align with the dimensions of $\bm{y}$, a reshape operation is applied to $\bm{e}w \in \mathbb{R}^{n \times m}$ to convert its dimension into $\mathbb{R}^{nm \times 1}$ (i.e., $\mathbb{R}^{t \times 1}$), which is omitted in the equation for simplicity.
To obtain the position of the predicted waveform, we will use non-maximum suppression (NMS) to eliminate the redundant output waveform probability. When the waveform probability reaches the set threshold at the $i$-th position, it is considered that there is a waveform. The formula is as follows:
\begin{gather}
    \hat{\bm{y}}=\{i|\operatorname{NMS}(\sigma(\bm{e}w))_i>\epsilon\}, \tag{d}
\end{gather}
where $\operatorname{NMS}$ operator is the non-maximum suppression algorithm, as shown in Algorithm \ref{alg:nms}, the parameter $\tau$ is setting 19~\cite{van1993heart}. $\operatorname{NMS}(\sigma(\bm{e}w))_i$ is the waveform probability at the $i$-th position, $\epsilon$ is a threshold selected by Youden's J index~\cite{youden1950index}.

\noindent\textbf{Generation}. 
For the denoise subtask, we enhance MITDB~\cite{MITDB} and PTBXL~\cite{PTBXL} by adding noise from MIT-BIH Noise Stress Test Database\cite{moody1984noise}:
\begin{gather}
\tilde{\bm{x}}=\bm{x}+\sqrt{P_x/(P_n10^{\gamma/10})}\cdot\bm{n}, \tag{e}
\end{gather}
where $\tilde{\bm{x}}$ is the noise ECG after origin $\bm{x}$ is added with noise $\bm{n}$, $P_x=\frac{1}{t}\|\bm{x}\|^2$ is the power of $\bm{x}$ signal, $P_n=\frac{1}{t}\|\bm{n}\|^2$ is the power of $\bm{n}$ noise. $\gamma$ is the target signal-to-noise ratio.

\begin{algorithm}[!ht]
\caption{Non-Maximum Suppression (NMS)}
\label{alg:nms}
\begin{algorithmic}
\State \textbf{Input:} Waveform Probability $S \in \mathbb{R}^t$; 
 Threshold $\tau$  
\State \textbf{Output:} Retained Waveform Probability $S_{\text{retain}}$ 
\State \textbf{begin}
\State $I \gets \text{argsort}(S, \text{descending})$ 
\State $I \gets I[S[I] > 0]$ \Comment{Exclude zero scores}

\State $I_{\text{retain}} \gets \emptyset$ 
\While{$I \neq \emptyset$}
    \State $I_{\text{retain}} \gets I_{\text{retain}} \cup \{I[0]\}$ \Comment{Select top candidate}
    \State $I \gets I[|I - I[0]| > \tau]$ \Comment{Suppress $\tau$-neighbors}
\EndWhile
\State \Return $S_{\text{retain}}=S[I_{\text{retain}}]$
\end{algorithmic}
\end{algorithm}

The MIT-BIH Noise Stress Test Database contains three types of noise: baseline wander, muscle artefact, and electrode motion artefact. Referencing previous work \cite{chatterjee2020review}, we mainly use the electrode motion artefact noise to expand the dataset, with a total of $\{-6, 0, 6, 12, 18, 24\}$ 6 types of signal-to-noise ratio.

\noindent\textbf{Feature-based Fréchet Distance}. 
We re-write the FFD as follows:
\begin{gather}
\label{eq:ffd_app}
\scalebox{1}{$\text{FFD}(e, \hat{e}) = \frac{1}{\sqrt{k}} \left( \left\| \mu - \hat{\mu} \right\|^2 + \text{Tr}\left( \Sigma + \hat{\Sigma} - 2 (\Sigma \hat{\Sigma})^{\frac{1}{2}} \right) \right)$},\tag{f} 
\end{gather}  
where $\text{Tr}$ is the trace operator, $\mu$ and $\Sigma$ denote the mean vector and covariance matrix of real ECG features, while $\hat{\mu}$ and $\hat{\Sigma}$ denote those of generated ECG features.

To calculate Eq.~\eqref{eq:ffd_app}, empirically, $\mu,\Sigma$ could be estimated by  $\tilde{\mu},\tilde{\Sigma}$ as follows:
\begin{gather}
\label{eq:mu_app}
\scalebox{1}{$ \tilde{\mu} = \frac{1}{N}\sum_{i=1}^N \bm{e}_i, \quad
 \tilde{\Sigma} = \frac{1}{N-1}\sum_{i=1}^N (\bm{e}_i - \tilde{\mu})^\top(\bm{e}_i - \tilde{\mu})$
    },\tag{g}
\end{gather} 
where $N$ denotes the number of real ECG samples, $\bm{e}_i$ represents the feature of $\bm{x}_i$ mapped by $f$, which could be implemented as a standard time-series feature extractor and the details are given in the experiments. $\hat{\mu}$ and $\hat{\Sigma}$ could be estimated in the same way.

\begin{figure}[htbp]
    \centering
    \includegraphics[width=0.8\linewidth]{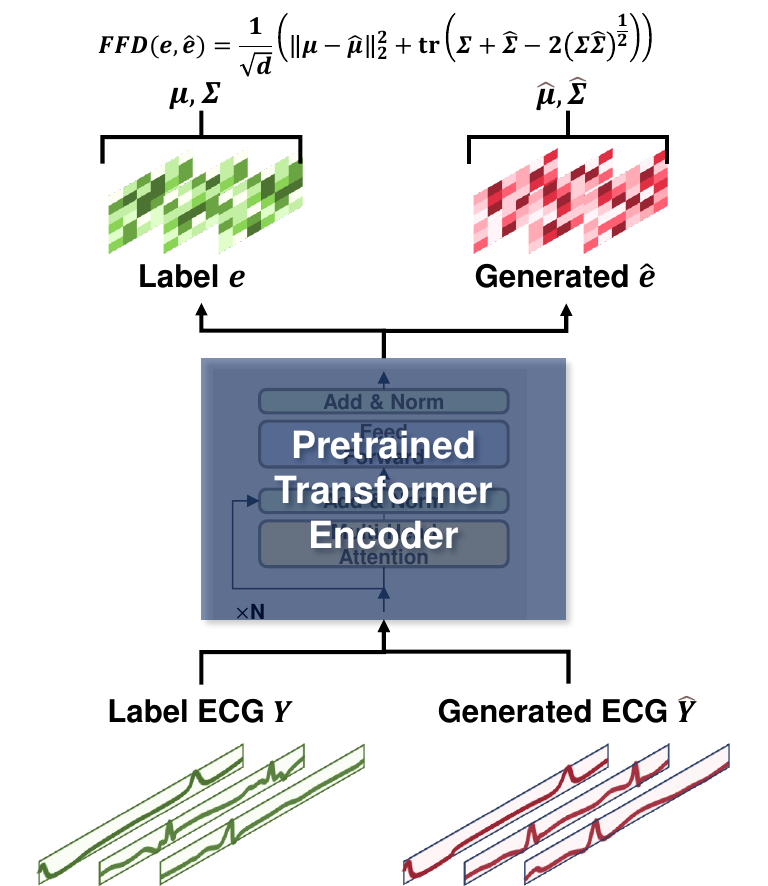}
  \caption{Feature-based Fréchet Distance calculation pipeline. The green part is the ground truth ECG and features, and the red is the generated ECG and features.}
  \label{fig:ffd_app}
  % \vspace{-8pt}
\end{figure}

As shown in Figure~\ref{fig:ffd_app}, the ground truth and generated ECG are input into a mapping $f$ to calculate the features, and FFD is calculated using Eq.~\eqref{eq:ffd_app} and Eq.~\eqref{eq:mu_app} in the main paper. Our mapping $f$ is a pre-trained transformer encoder that uses a mask token prediction object. The ECG signal $\bm{x}$ of length $t$ is patched into $n$ tokens $\bm{s}=\{s_i\}^n_{i=1}$, the length of each token is $m=t/n$, and the model predicts $k$ mask token $\{s_i\}^k$ set to 0. The loss is as follows:
\begin{gather}
\label{eq:mtp_app}
    \mathcal{L}_{\text{MTP}}=\frac{1}{k}\sum_i\frac{1}{m}\|s_i-e_iw\|^2, \tag{h}
\end{gather}
where $s_i$ is the ground truth of $i$-th masked token, \(e_i\!\in\! \mathbb{R}^{1 \times d}\) denotes the features of time $i$ extracted by model, $d$ denotes the hidden dimension, and \(w \in \mathbb{R}^{d \times m}\) is a linear projector.

While training, the number of masked tokens $n_m$ obeys the following probability:
\begin{gather}
% \label{eq:mtp_app}
    n_{m}=\left\{\begin{matrix} 0.5n&p\le0.5,\\\mathcal{U}(n/4,3n/4)&p>0.5,
\end{matrix}\right. \tag{i}
\end{gather}
where $n$ is the number of tokens $\bm{s}$, and $p$ is uniformly distributed from 0 to 1, i.e., $p\sim\mathcal{U}(0,1)$. There is a half probability that the masked tokens are half of the total tokens and the remaining half probability that the number of masked tokens obeys a uniform distribution from $n/4$ to $3n/4$, i.e., $n_m\sim\mathcal{U}(n/4,3n/4)$.

\subsection*{Dataset}

\begin{table*}[htbp]
        \renewcommand{\arraystretch}{1.}
    \scalebox{1}{
\begin{tabularx}{1\textwidth}{l|l|lXXXXX}
\toprule
\multicolumn{1}{c|}{\textbf{Task}} & \multicolumn{1}{c|}{\textbf{Name}} & \multicolumn{1}{c}{\textbf{Describe}} & \multicolumn{1}{c}{\textbf{\# Sample}} & \multicolumn{1}{c}{\textbf{\# Subject}} & \multicolumn{1}{c}{\textbf{Channel}} & \multicolumn{1}{c}{\textbf{Frequency}} & \multicolumn{1}{c}{\textbf{Size}} \\
\hline
\multirow{6}{*}{\textbf{Classification}} & \textbf{CPSC2018} \cite{CPSC2018} & 7     & 205111 & 5847  & 12    & 500Hz & 1.23GB \\
      & \textbf{CPSC2021} \cite{CPSC2021} & 3     & 330032 & 785   & 2     & 200Hz & 1.29GB \\
      & \textbf{AF} \cite{AF} & 2     & 8400  & 35    & 7     & 125Hz & 0.02GB \\
      & \textbf{NIFEADB} \cite{NIFEADB} & 2     & 21293 & 26    & 6     & 1000Hz & 0.17GB \\
      & \textbf{RFAA} \cite{RFAA} & 4     & 38640 & 315   & 4     & 250Hz & 0.40GB \\
      & \textbf{SPB} \cite{goldberger2000physiobank} & 6     & 34560 & 8     & 12    & 257Hz & 0.76GB \\
\hline
\multirow{6}{*}{\textbf{Detection}} & \textbf{FEPL} \cite{FEPL} & Fetal QRS & 37080 & 20    & 10    & 500Hz & 0.12GB \\
      & \textbf{CPSC2020} \cite{CPSC2020} & T-wave & 175782 & 10    & 1     & 400Hz & 0.49GB \\
      & \textbf{MITDB} \cite{MITDB} & QRS   & 20810 & 71    & 2     & 360Hz & 0.11GB \\
      & \textbf{MITPDB} \cite{MITPDB} & P-wave & 4269  & 12    & 2     & 360Hz & 0.02GB \\
      & \textbf{NFE} \cite{NFE} & Fetal QRS & 1200  & 25    & 4     & 1000Hz & 0.01GB \\
      & \textbf{SVDB} \cite{SVDB} & QRS   & 56160 & 78    & 2     & 128Hz & 0.05GB \\
\hline
\multirow{6}{*}{\textbf{Forecast}} & \textbf{CPSC2019} \cite{CPSC2019} &       & 2000  & 2000  & 1     & 500Hz & 0.02GB \\
      & \textbf{RDBH} \cite{RDBH} &       & 22200 & 198   & 7     & 360Hz & 0.36GB \\
      & \textbf{MIMICSub} \cite{MIMIC3Sub} &       & 5552708 & 1210    & 8     & 125Hz & 11.47GB \\
      & \textbf{NFE} \cite{NFE} &       & 5467  & 125   & 4     & 1000Hz & 0.05GB \\
      & \textbf{DALIA} \cite{DALIA} &  & 25887  & 15  & 2  & \multicolumn{1}{l}{100Hz} & \multicolumn{1}{l}{0.09GB} \\
      & \textbf{PTB} \cite{PTB} &       & 130455 & 516   & 12    & 1000Hz & 1.27GB \\
\hline
\multirow{6}{*}{\textbf{Generation}} & \textbf{ADFECGDB} \cite{ADFECGDB} & Maternal ECG & 1200  & 5     & 5     & 1000Hz & 0.01GB \\
      & \textbf{FEPL} \cite{FEPL} & Maternal ECG & 7800  & 10    & 10    & 500Hz & 0.05GB \\
      & \textbf{BIDMC} \cite{BIDMC} & PPG   & 4992  & 52    & 7     & 125Hz & 0.20GB \\
      & \textbf{MITDB} \cite{MITDB} & Noise ECG & 116196 & 69    & 2     & 360Hz & 0.11GB \\
      & \textbf{PTBXL} \cite{PTBXL} & Noise ECG & 262044 & 21837 & 12    & 500Hz & 2.45GB \\
      & \textbf{SST} \cite{SST} & PCG   & 6760  & 338   & 9     & 1000Hz & 0.17GB \\
\bottomrule
\end{tabularx}%
}
\caption{Summary of scale information for each dataset. All columns except for "Frequency" represent post-preprocessed values. The same dataset may have different information for different tasks.}
\label{tab:datasets_app}
\end{table*}

Next, we will detail the various datasets mentioned in the main paper. More information on these datasets can be found in Table \ref{tab:datasets_app}.

\noindent\textbf{Classification.}
We collected 6 datasets for classification:

\begin{itemize}
    \item The China Physiological Signal Challenge 2018 (CPSC2018) 
    % comprises ECG signals of one normal type and eight abnormal types, 
    contains a diverse set of cardiac conditions. After processing, there are 7 classes.
    
    \item The China Physiological Signal Challenge 2021 (CPSC2021) is recorded from 12-lead Holter or 3-lead wearable ECG monitoring devices. Challenge focuses on atrial fibrillation diagnosis and has 3 classes in total.
    
    \item AF Classification (AF)~\cite{AF} contains short-term ECG recordings with binary class for atrial fibrillation.
    
    \item Non-Invasive Fetal ECG Arrhythmia (NIFEADB)~\cite{NIFEADB} contains four or five abdominal channels and one thoracic maternal channel. It is required to diagnose fetal arrhythmias from these ECGs

    \item Reducing False Arrhythmia Alarms (RFAA)~\cite{RFAA} uses bedside monitor data for 1,250 life-threatening arrhythmia alarms recorded by the bedside monitor. 
    % Expert annotators reviewed each alarm and labeled it as ``true'', ``false'', or ``cannot tell''. 
    We processed it into 4 classes.

    \item St. Petersburg Arrhythmia Database~(SPB)~\cite{goldberger2000physiobank} contains ECG recordings from patients undergoing coronary artery disease (17 men and 15 women, aged 18–80 years;). It comprises 75 ECG recordings annotated with 6 diseases.
\end{itemize}

\noindent\textbf{Detection.}
We evaluate 6 datasets for detection:

\begin{itemize} 
    \item Fetal ECG with Pregnancy and Labor~(FEPL)~\cite{FEPL} contains multiple records. It includes the abdominal and FECG signals, 
    % as well as a text file with information about the positions of the maternal and fetal QRS complexes, 
    with reliability markers for the fetal R waves.

    \item The China Physiological Signal Challenge 2020 (CPSC2020) consists of 10 single-lead ECG records collected from patients with arrhythmias, each lasting approximately 24 hours. It was collected from arrhythmia patients with abnormal T-wave annotations.

    \item MIT-BIH Arrhythmia Database~({MITDB}, {MITPDB})~\cite{MITDB,MITPDB}, MITDB contains 48 ECG records and annotates the QRS complex; MITPDB complements MITDB by adding P waves annotations.

    \item Noninvasive Fetal ECG~(NFE)~\cite{NFE} consists of some fetal ECG recordings. Each recording includes 4 non-invasive abdominal signals annotated with fetal R waves.

    \item Supraventricular Arrhythmia Database (SVDB)~\cite{SVDB} contains 78 half-hour ECG recordings, which supplement the cases of supraventricular arrhythmias in the MITDB, annotated with QRS complexes.
\end{itemize}

\noindent\textbf{Forecasting.} We selected 6 datasets for forecasting:
\begin{itemize}
    \item The China Physiological Signal Challenge 2019 (CPSC2019)~\cite{CPSC2019} includes 2,000 single-lead ECG recordings collected from patients with cardiovascular disease. It is often used in ECG detection tasks.
    \item PPG-DaLiA (DALIA)~\cite{DALIA} contains 
    % physiological and motion data recorded from wrist- and chest-worn devices from 15 subjects while performing various activities. 
    Contains ECG, photoplethysmography, and 3D accelerometer data while compensating for motion artefacts. It is often used in ECG generation tasks.
    \item PTB Diagnostic ECG Database (PTB, PTBXL)~\cite{PTB,PTBXL}, PTB contains 549 records from 290 subjects. Each record includes 15 simultaneously measured signals: the conventional 12-lead and 3 Frank-lead ECGs. PTBXL ECG contains 21,799 clinical 12-lead ECGs from 18,869 patients, with each ECG lasting 10 seconds. They are often used in ECG classification tasks.
    \item A subset of MIMIC-III Waveform Database Matched Subset~(MIMICSub)~\cite{MIMIC3Sub}, the original dataset contains 22,247 numerical records of 10,282 different ICU patients. We selected the first ECG recording of each patient in the folders ``p00'' and ``p03''.
    \item Detection of Heart Beats Dataset~(RDBH)~\cite{RDBH} contains 10-minute recordings. 
    Each recording contains four to eight signals. 
    % ; the first is the ECG signal, and the others are various physiological signals recorded simultaneously. 
    It is often used in ECG detection tasks.
    \item European ST-T Database~(EBD)~\cite{EBD} contains 90 annotated Holter recordings from 79 subjects. It is often used in ECG detection tasks.
\end{itemize}

\begin{table*}[ht]
\centering
   \renewcommand{\arraystretch}{1.}
    \scalebox{1}{
\begin{tabularx}{1.\textwidth}{l|>{\centering\arraybackslash}X>{\centering\arraybackslash}X>{\centering\arraybackslash}X>{\centering\arraybackslash}X>{\centering\arraybackslash}X>{\centering\arraybackslash}X>{\centering\arraybackslash}X}
\toprule
\multicolumn{1}{c|}{\multirow{2}[2]{*}{\textbf{Datasets}}} & \textbf{Informer} & \textbf{Medformer} & \textbf{UniTS} & \textbf{UniTS*} & \textbf{Timer} & \textbf{Timer*} & \textbf{PSSM} \\
      &  \cite{zhou2021informer}     & \cite{wang2024medformer}      &  \cite{gao2024units}     &      \cite{gao2024units}       &    \cite{liutimer}   &  \cite{liutimer}     & (Ours) \\
\hline
\textbf{CPSC2019} \cite{CPSC2019} & 1.021 & {1.013} & \underline{1.006} & 1.128 & 1.030 & 1.035 & \textbf{0.186} \\
\textbf{DALIA} \cite{DALIA} & 1.050 & \textbf{0.904} & 1.014 & 1.113 & 0.976 & \underline{0.919} & 0.956 \\
\textbf{RDBH} \cite{RDBH} & 0.850 & \textbf{0.802} & 0.984 & 1.063 & 0.830 & 0.869 & \underline{0.829} \\
\textbf{MIMICSub} \cite{MIMIC3Sub} & 0.601 & \textbf{0.522} & 0.961 & 1.050 & 0.819 & 0.594 & \underline{0.582} \\
\textbf{NFE} \cite{NFE} & 0.686 & \underline{0.577} & 0.945 & 1.036 & 0.686 & 0.968 & \textbf{0.345} \\
\textbf{PTB} \cite{PTB} & 0.376 & \underline{0.233} & 0.414 & 0.448 & 0.302 & 0.295 & \textbf{0.131} \\
\hline
\textbf{Average} & 0.764 &\underline{ 0.675} & 0.887 & 0.973 & 0.774 & 0.780 & \textbf{0.505} \\
\bottomrule
\end{tabularx}%

}
\caption{Test MSE on ECG forecasting datasets. A smaller MSE is better. Bold numbers indicate the best performance, underlined numbers indicate the runner-up, and ``*'' indicates that the model is further pre-trained on the ECG datasets. ECGPT~\cite{davies2024interpretable} is not applicable for this setting, and its performance is omitted.}
\label{tab:forecast_app}
\end{table*}

\begin{table*}[htbp]
  \centering

        \renewcommand{\arraystretch}{1.}
    \scalebox{1}{
\begin{tabularx}{1\textwidth}{l|>{\centering\arraybackslash}X>{\centering\arraybackslash}X>{\centering\arraybackslash}X>{\centering\arraybackslash}X>{\centering\arraybackslash}X>{\centering\arraybackslash}X>{\centering\arraybackslash}X>{\centering\arraybackslash}X}
\toprule
\multicolumn{1}{c|}{\multirow{2}[2]{*}{\textbf{Datasets}}} & \textbf{Informer} & \textbf{Medformer} & \textbf{UniTS} & \textbf{UniTS*} & \textbf{ECGPT} & \textbf{Timer} & \textbf{Timer*} & \textbf{PSSM} \\
      &  \cite{zhou2021informer}     & \cite{wang2024medformer}      &  \cite{gao2024units}     &      \cite{gao2024units} &  \cite{davies2024interpretable}     &    \cite{liutimer}   &  \cite{liutimer}     & (Ours)  \\
\hline
\textbf{MITDB} \cite{MITDB} & 0.208  & \underline{0.169}  & 0.321  & 0.297  & 0.346  & 0.232  & 0.187  & \textbf{0.082 } \\
\textbf{PTBXL} \cite{PTBXL} & 0.203  & 0.190  & 0.228  & 0.216  & 0.279  & 0.213  & \underline{0.161}  & \textbf{0.095 } \\
\textbf{ADFECGDB} \cite{ADFECGDB} & 0.905  & 0.912  & 0.891  & 0.900  & 0.902  & 0.872  & \underline{0.868}  & \textbf{0.587 } \\
\textbf{FEPL} \cite{FEPL} & 0.647  & 0.763  & 0.670  & 0.737  & 0.764  & 0.714  & \underline{0.519}  & \textbf{0.237 } \\
\textbf{BIDMC} \cite{BIDMC} & 0.985  & 0.984  & 0.850  & 0.906  & 1.023  & 0.809  & \underline{0.793}  & \textbf{0.483 } \\
\textbf{SST} \cite{SST} & 0.925  & 0.928  & \underline{0.916}  & 0.930  & 0.937  & 0.961  & 1.010  & \textbf{0.645 } \\
\hline
\textbf{Average} & 0.646  & 0.658  & 0.646  & 0.664  & 0.708  & 0.634  & \underline{0.589}  & \textbf{0.355 } \\
\bottomrule
\end{tabularx}%

}
\caption{Test MSE on ECG generation datasets. A smaller MSE is better. Bold numbers indicate the best performance, underlined numbers indicate the runner-up, and ``*'' indicates that the model is further pre-trained on the ECG datasets.}
  \label{tab:generation_app}%
\end{table*}%

\noindent\textbf{Generation.}
Generation evaluation contains 6 datasets, including FEPL. For the denoising subtask, we augment the MITDB and PTBXL\cite{PTBXL} by injecting noise from the MIT-BIH Noise Stress Test Database \cite{moody1984noise}. The remaining datasets are described below:

\begin{itemize}
    \item Abdominal and Direct Fetal ECG (ADFECGDB)~\cite{ADFECGDB} contains maternal abdominal ECG and direct fetal ECG obtained from 5 pregnant subjects.

    \item BIDMC PPG and Respiration Dataset (BIDMC)~\cite{BIDMC} extracts ECG from the MIMIC II Waveform Database~\cite{saeed2011multiparameter}, contains 53 records in the dataset, each lasting 8 minutes. Two annotators manually annotated individual respirations in each record using impedance respiration signals. Contains physiological signals such as PPG, impedance respiration signals, and ECG.

    \item SensSmartTech Database (SST)~\cite{SST} contains 338 30-second polycardiogram signals, each containing 10 channels: 4 ECG, 4 PPG, 1 PCG, and 1 ACC channels. These signals come from 32 subjects (18 females and 14 males).
\end{itemize}

We further organized an extension ECG Dataset for LTMs pretraining, comprising AF, CPSC2018, SPB, PTB, PTBXL, NFE, MIMICSub, The Georgia 12-lead ECG Challenge Database~\cite{goldberger2000physiobank}, and the Shaoxing and Ningbo Hospital ECG Database~\cite{Shaoxing, Ningbo}. This dataset contains 98,513 subjects (24.59GB raw data) processed into 8,414,834 training samples. Although the extension ECG Dataset overlaps with other datasets, the training objectives differ (e.g., AF and CPSC2018), or the overlapping samples are removed (e.g., MIMICSub).
\begin{itemize}
    \item The Georgia 12-lead ECG Challenge Database~\cite{goldberger2000physiobank} contains 20,672 ECGs. Each recording is between 5 and 10 seconds long, with a sampling frequency of 500 Hz.
    \item The Shaoxing and Ningbo Hospital ECG Database~\cite{Shaoxing,Ningbo} contains 45,152 10-second 12-lead ECGs.
\end{itemize}

\subsection*{Implementation Details}
We evaluated advanced LTMs: Timer~\cite{liutimer}, UniTS~\cite{gao2024units}, and ECGPT~\cite{davies2024interpretable}. Both raw versions and variants are pre-trained on the extension ECG Dataset, denoted by ``*''. The mapping $f$ of FFD is the LTM constructed by the transformer encoder. Additionally, we tested state-of-the-art transformer-based end-to-end models: Informer~\cite{zhou2021informer} with sparse attention mechanisms and Medformer~\cite{wang2024medformer} featuring multi-scale data fusion alongside our proposed PSSM.

We have already mentioned the general settings in the main paper. Here are the settings for each model:
\begin{itemize}

    \item Informer~\cite{zhou2021informer}: We use the encoder part of Informer, which consists of 3 layers with the hidden layer dimension set to 128.
    
    \item Medformer~\cite{wang2024medformer}: We use the encoder part of Medformer, which consists of 4 layers with the hidden layer dimension set to 128. The model-specific path length list is $\{8,8,8,$ $16,16,16\}$.

    \item UniTS~\cite{gao2024units}: Raw UniTS remains the same as the original paper, i.e., 3-layer encoder, hidden layer dimension is 64, patch length is 16. The patch length of UniTS* is 50, and the "task data config" is adjusted to the information of the extension ECG Dataset.
    
    \item Timer~\cite{liutimer}: Raw Timer is the same as the original paper, i.e., 8-layer transformer decoder, hidden layer dimension is 1024, patch length is 96. Timer* patch length is 50.
    
    \item PSSM: The PSSM encoder and decoder have 4 layers, i.e., $l=4$ and $d=32$ except for the forecasting where $d=256$ is trained on the extension ECG Dataset to ensure fairness in dataset scale and model capacity.

    \item FFD mapping $f$: It consists of 8 layers of transformer encoders, with a hidden layer dimension 1024. 
\end{itemize}

\subsection*{Further Results and Analysis}
% \noindent\textbf{MSE Results}.

\noindent\textbf{MSE Results}.
(a) The MSE results of forecasting are shown in Table \ref{tab:forecast_app}. The best LTMs performer is Timer* with an MSE of 0.774, and the best model performer is PSSM with an MSE of 0.505, which is 25.2\% higher than Medformer and 34.7\% higher than Timer. (b) The MSE results of generation are shown in Table \ref{tab:generation_app}. The best LTMs performer is Timer* with an MSE of 0.589, and the model performer is PSSM with an MSE of 0.355, which is 45.1\% higher than Informer and 39.8\% higher than Timer*.

Although the ``*'' models outperform their raw versions (e.g., Timer$^*$ reduces MSE by 7.1\% compared to Timer), they still fall short of PSSM. Limited fine-tuning parameters, while necessary to preserve pretraining benefits, may partially explain this gap. In forecasting without fine-tuning, Timer$^*$ also weakly performs as well as PSSM, with the MSE up 54.5\%.

\begin{figure}[htbp]
    \centering
    \includegraphics[width=\linewidth]{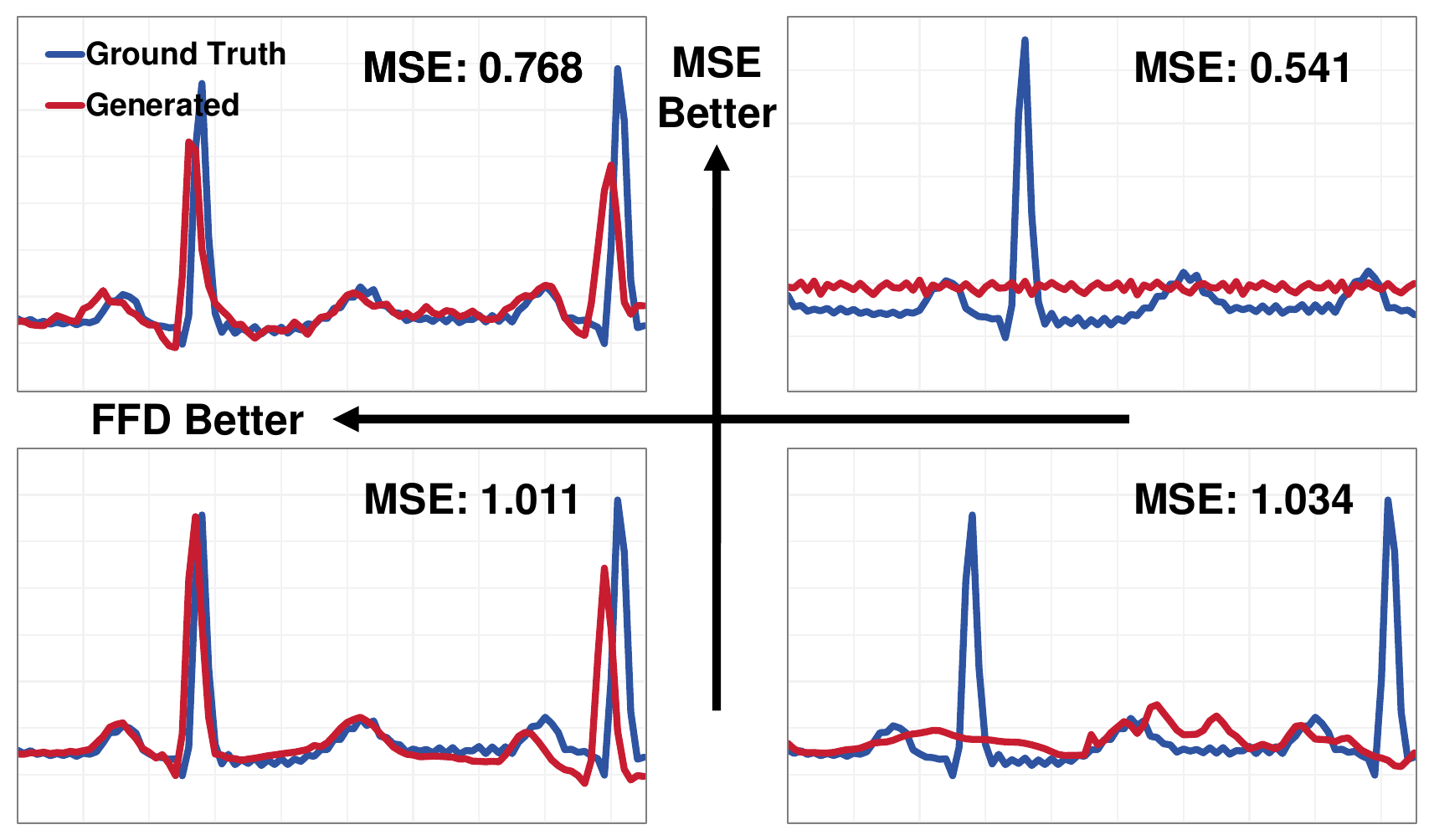}
    \caption{Qualitative results are MSE and FFD differences. Blue curves are the ground truth ECG, and red curves are the generated ECG. The MSE is marked in the upper right corner of each Quadrant. The data comes from the NFE dataset\cite{NFE} in forecasting.}
    \label{fig:mse_vs_ffd_app}
    \vspace{-8pt}
\end{figure}
\noindent\textbf{MSE vs. FFD}.
\textit{The limitation of MSE} is its inability to accurately assess the quality of the generated ECG, i.e., the generated high quality may have a higher MSE. As illustrated in Figure~\ref{fig:mse_vs_ffd_app}, flatline signals in Quadrant I achieve smaller MSE values than those in Quadrant III yet exhibit significantly worse ECG quality. This discrepancy arises from ECG’s characteristic extremum values, e.g., minima at P-wave and maxima at R-wave. Minor temporal shifts in generated ECGs can cause misalignment between generated maxima and true minima, artificially inflating MSE. Thus, MSE alone is misleading.

\textit{The advantage of FFD} is the ability to accurately evaluate the generated ECGs semantics, i.e., the lower the FFD, the higher the quality of the generated ECGs. In addition to the discussion in Section 4.3 of the main paper as evidence, as shown in Figure~\ref{fig:mse_vs_ffd_app}, generated ECGs in Quadrants II and III with better FFD retain diagnostically valid patterns, whereas those in Quadrants I and IV with worse FFD show degraded ECG patterns.

In summary, FFD is a critical complement to better evaluate ECG generation quality. While MSE is limited to assessing point-wise signal similarity, FFD captures higher-order feature distributions, enabling a more comprehensive structural and diagnostic fidelity evaluation.

\begin{figure}
    \centering
    \includegraphics[width=\linewidth]{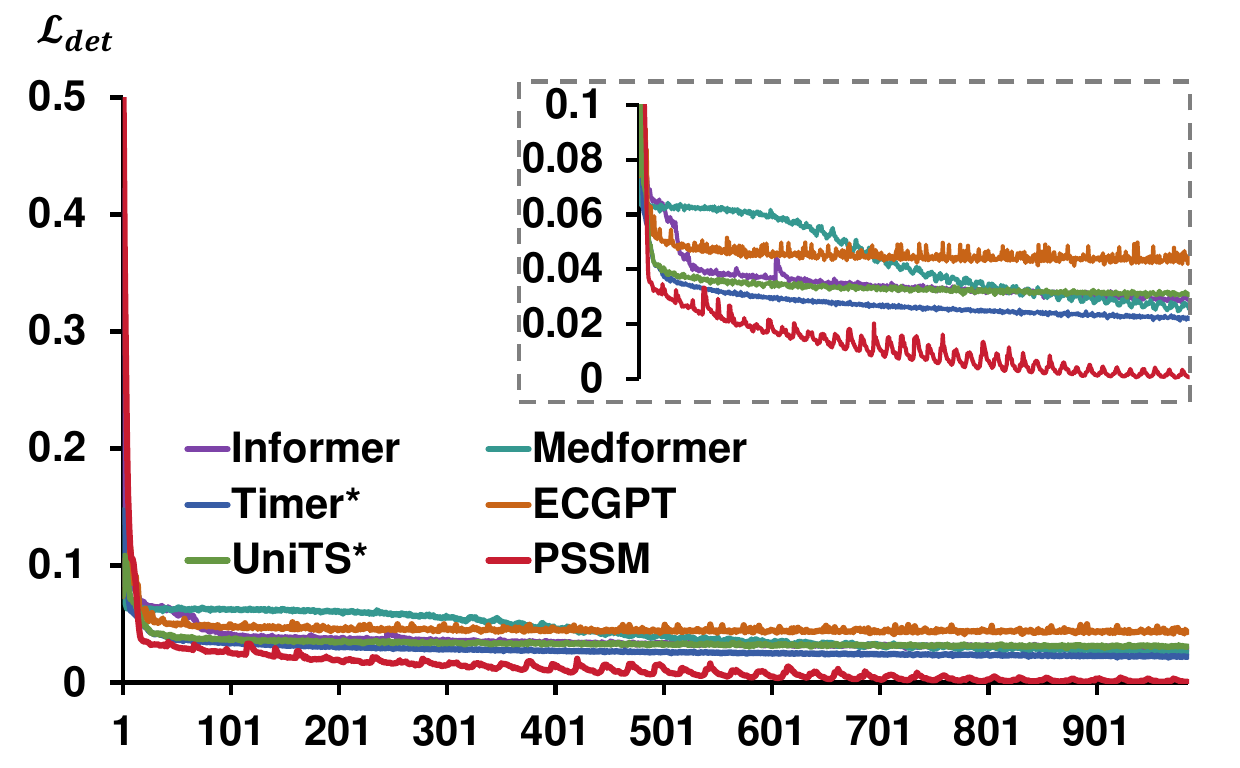}
    \caption{Visualization of training losses in detection across 6 architectures: Informer~\cite{zhou2021informer}, Medformer~\cite{wang2024medformer}, UniTS*~\cite{gao2024units}, ECGPT~\cite{davies2024interpretable}, Timer*~\cite{liutimer}, and PSSM. Derived from SVDB dataset~\cite{SVDB}.}
    \label{fig:detection_loss_app}
    \vspace{-8pt}
\end{figure}
\noindent\textbf{Detection Further Analysis}.
We re-show the detection results as shown in Table~\ref{tab:detection}. The performance gap between the best model PSSM and the weakest model Medformer is significant, with an F1-score difference of 0.494 (151.2\%). While all models converged during training, as shown in Figure~\ref{fig:detection_loss_app}, with training losses reduced below 0.1, only Timer* and PSSM achieved losses below 0.04. These two models exhibit superior F1-scores in Table~\ref{tab:detection}.

This discrepancy stems from the severe class imbalance in detection (waveform annotations positions <5\% in an ECGs recording $\bm{x}$). Models tend to lazily predict all positions as no waveform, rapidly driving training loss below 0.1. However, high-performing models (e.g., Timer*, PSSM) further refine their ability to distinguish the positive class, reducing losses below 0.04. Consequently, even minor differences in training loss correspond to substantial gaps in test F1-scores.

\end{document}